\documentclass[12pt,preprint]{aastex}
\usepackage{natbib,graphicx,latexsym}
\newcommand{\lsim}{\
\raise-2.truept\hbox{\rlap{\hbox{$\sim$}}\raise5.truept\hbox{$<$}\ }}
\newcommand{\gsim}{\
\raise-2.truept\hbox{\rlap{\hbox{$\sim$}}\raise5.truept\hbox{$>$}\ }}

\begin{document}

\title{The globular cluster system in NGC\,5866: optical observations
from HST Advanced Camera for Surveys\altaffilmark{1}}

\author{Cantiello, Michele\altaffilmark{2,3}}
\author{Blakeslee, John P.\altaffilmark{2}}
\author{Raimondo, Gabriella\altaffilmark{3}}

\altaffiltext{1}{Based on observations made with the NASA/ESA Hubble
Space Telescope, obtained from the Data Archive at the Space
Telescope Science Institute, which is operated by the Association of
Universities for Research in Astronomy, Inc., under NASA contract NAS
5-26555.}
\altaffiltext{2}{Department of Physics and Astronomy, Washington State University,
Pullman, WA 99164;  email: cantiello, jblakes@wsu.edu}
\altaffiltext{3}{INAF--Osservatorio Astronomico di
Teramo, Via M. Maggini, I-64100 Teramo, Italy; email: cantiello, raimondo@oa-teramo.inaf.it}

\begin{abstract}
We perform a detailed study of the Globular Cluster (GC) system in the
galaxy NGC\,5866 based on F435W, F555W, and F625W ($\sim$ B, V, and R)
HST Advanced Camera for Surveys images.

Adopting color, size and shape selection criteria, the final list of
GC candidates comprises 109 objects, with small estimated
contamination from background galaxies, and foreground stars.

The color distribution of the final GC sample has a bimodal
form. Adopting color to metallicity transformations derived from the
Teramo--SPoT simple stellar population model, we estimate a
metallicity [Fe/H]$\sim -1.5$, and $-$0.6 dex for the blue and red
peaks, respectively. A similar result is found if the empirical
color-metallicity relations derived from Galactic GCs data are used.

The two subpopulations show some of the features commonly observed in
the GC system of other galaxies, like a ``blue tilt'',
higher central concentrations of the red subsystem, and larger
half--light radii at larger galactocentric distances. However, we do
not find evidence of a substantial difference between the average sizes of
red and blue clusters.

Our analysis of the GC Luminosity Function indicates a V-band
Turn-Over Magnitude V$_0^{TOM}$=23.46$\pm$0.06, or
M$_{V,0}^{TOM}\sim-7.29 \pm 0.10$ mag, using the distance modulus
derived from the average of SBF and the PNLF distances. The absolute
Turn-Over Magnitude obtained agrees well with calibrations from
literature. The specific frequency is measured to be $S_N=1.4 \pm
0.3$, typical for galaxies of this type.

\end{abstract}


\section{Introduction}

The study of the Globular Cluster (GC) system in galaxies is one of
the fundamental keystones to understand the physical processes at the
base of the formation and evolution of galaxies. The investigation of
GC system properties involves aspects of several different fields in
astronomy: stellar evolution, hierarchical assembly, the determination
of extra--galactic distances, etc.

The first applications of the GCs as an investigation tool came in the
earliest years of extra--galactic astronomy.  Single bright GC were
recognized as possible distance indicator in the 1950s \citep{baum55},
but it took few decades to develop the technique as it is applied
now. The systematic study of the GC Luminosity Function (GCLF), in
fact, has shown that it has a universal nearly gaussian shape. It is
now recognized that the Turn Over Magnitude (TOM) of the GCLF is a
reliable distance indicator for most galaxies, although some
exceptions exist \citep{f00,richtler03,jordan07}.

Another, more recent discovery in the field of GC astronomy, is that
the presence of a color bimodality is quite common in galaxies. The
color bimodality has been mostly interpreted as a bimodal metallicity
distribution \citep{harris01,brodie06}, leading to develop several
possible GC formation scenarios, useful to constrain the history of
galaxy formation. More recently, \citet{kundu07} using near-IR
photometry, have shown that the GC bimodality in the giant ellipticals
M\,87 and NGC\,4472 might represent true metallicity
subpopulations. However, whether or not the bimodal color distribution
is linked to the presence of two metallicity subpopulations is still a
matter of debate, and there are claims that the optical properties of
a bimodal GC system are not necessarily due to a bimodal metallicity
distribution. For example, it has been shown that a bimodal color
distributions might be generated by the combination of a unimodal
metallicity distribution and a non-linear color-metallicity relation
of the appropriate shape \citep{richtler06,yoon06}.

Furthermore, the advent of space-based astronomy has provided new
information in this field. Most of all, the high resolution optical
imaging of HST makes it possible to examine the shape properties,
i.e. size and ellipticity, for an increasing number of extra--galactic
GCs, up to distances as far as the Virgo Cluster \citep{jordan07} and
beyond. The examination of the characteristic sizes, and radial
distribution of the blue and red GC subpopulations, enables the study
of new phenomena related to the formation history of the GC system
\citep{larsen03,jordan04}.

In this paper we take advantage of archival HST images of the galaxy
NGC\,5866, taken with the Advanced Camera for Surveys (ACS) in the
Wide Filed Camera mode, to carry out the first detailed
multi-wavelength study of the GC system in this galaxy.

NGC\,5866 is a lenticular galaxy, with a sharp dust lane nearly edge
on. Being a disk galaxy, the GC system is expected to be less
populated with respect to ellipticals
\citep[e.g.][]{harris81,brodie06}. However, the quality of these ACS
data allows us to analyze in some detail the clusters properties.

The weighted average of distances moduli obtained using the SBF
\citep[][using the new zeropoint from Jensen et al, 2003]{tonry01},
and PNLF \citep{ciardullo02} methods, provide a distance modulus
$\mu_0=30.75 \pm 0.08$, or 14.1$\pm$0.5 Mpc. At this distance, the GCs
of the galaxy are near the ACS resolution limit, allowing us to carry out
the size and shape analysis of the GC system.

The analysis of GC properties will be presented in the next sections.
The observations and data reduction procedure are introduced in
Section 2. Section 3 deals with the GC photometry and selection
criteria. The color bimodality, color-magnitude relations, and the
spatial distribution of the total GC system, as well as the two GC
subpopulations, are discussed in Section 4. In the same Section we
also make a comparison with stellar synthesis models. Finally, a study
of GCLF and specific frequency is also presented in Section
4. We summarize and conclude the paper in Section 5.

\section{Data reduction}

The ACS/WFC camera observations of NGC\,5866 were taken in February
2006 with the F435W, F555W and F625W filters ($\sim$ B, V, and R). The
total exposure times for the three different passbands were: 3900s,
2800s and 2200s, respectively.

Standard reduction was carried out on raw images using the APSIS image
processing software \citep{blake03}. After cosmic--ray rejection,
alignment, ``drizzling'', and final image combination, we adopted an
iterative procedure to determine the sky background and the galaxy
best fit models, independently for the three passbands.

Since the galaxy fills the whole field of view of the ACS, the sky
background determined in the outer corners of the CCD image is not
accurate.  We obtain the sky values extrapolating from the fit of a de
Vaucouleurs $r^{1/4}$ profile.

Briefly, after masking out a few bright/saturated foreground stars, we
adopt a provisional sky value from the corner with the lowest median
number of counts. This value is subtracted to the image before
starting the galaxy fitting procedure. Then, we fit the galaxy
isophotes using the IRAF/STSDAS task ELLIPSE\footnote{IRAF is
distributed by the National Optical Astronomy Observatories, which are
operated by the Association of Universities for Research in Astronomy,
Inc., under cooperative agreement with the National Science
Foundation.}, which is based on the method described by
\citet{jedrzejewski87}.  Once the preliminary galaxy model is
subtracted from the sky--subtracted frame, a wealth of faint sources
appears together with the dust disk. A mask of these sources is
obtained using a sigma detection method, and combined with the
previous mask. The new mask is then fed to ELLIPSE to refit the
galaxy's isophotes.

After the geometric profile of the isophotes has been determined, we
measure the galaxy flux in elliptical isophotes. At last, to improve
the estimation of the sky, we fit the surface brightness profiles with
a de Vaucouleurs $r^{1/4}$ profile plus a constant sky offset.  The
sky offset fitted is then adopted as the new sky value, and the whole
procedure of galaxy fitting, source masking, surface brightness
profile fitting and sky estimation is repeated, until convergence.
This procedure is adopted for all three filters.

Once the sky and galaxy model have been subtracted from the frame,
some large--scale deviations are still present in the frame. We remove
these deviations using the background map obtained running SExtractor
\citep{bertin96} on the sky and galaxy subtracted frame. Part of the
procedure adopted here is the same developed by \citet{cantiello05} to
process ACS images for SBF analysis. We refer the reader to that paper
for more details on the whole procedure of sky+galaxy+large scale
residuals subtraction.

It is worth noting that in \citet{cantiello05} we have made some
checks on the stability of the sky estimation against the fitting
profile chosen. There, it was shown that, adopting a more general
Sersic $r^{1/n}$ profile, the sky values obtained agree within
uncertainties with those obtained adopting a de Vaucouleurs
profile. In addition, in \citet{cantiello07}, we compared the sky
values obtained applying our method with the sky obtained by
\citet{sikkema06} using a different approach, over a sample of 6
peculiar galaxies. As a result we found that the differences between
the two sky background determinations is on average $\lsim$3\%.

\begin{figure}
\epsscale{1.}
\plottwo{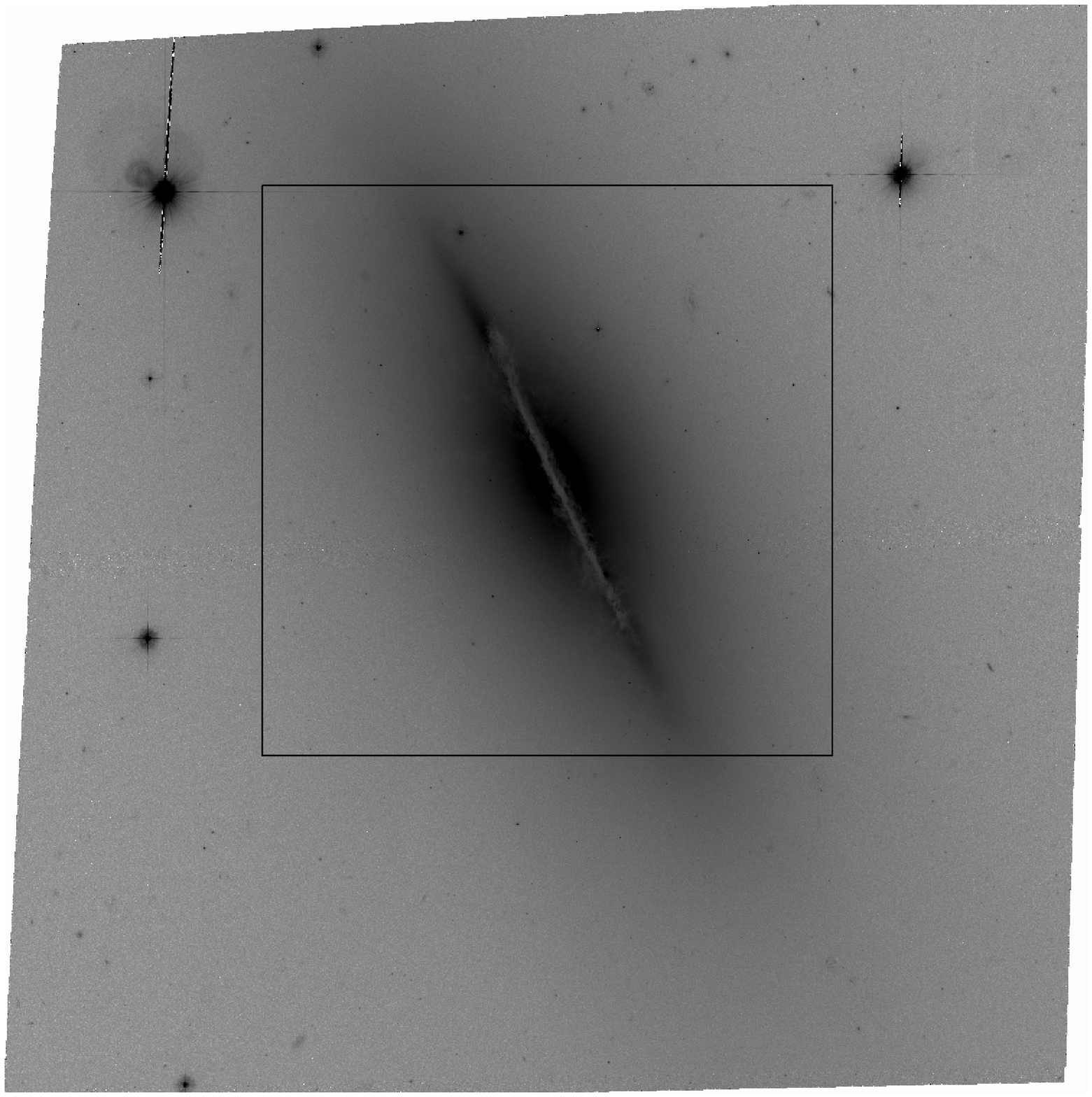}{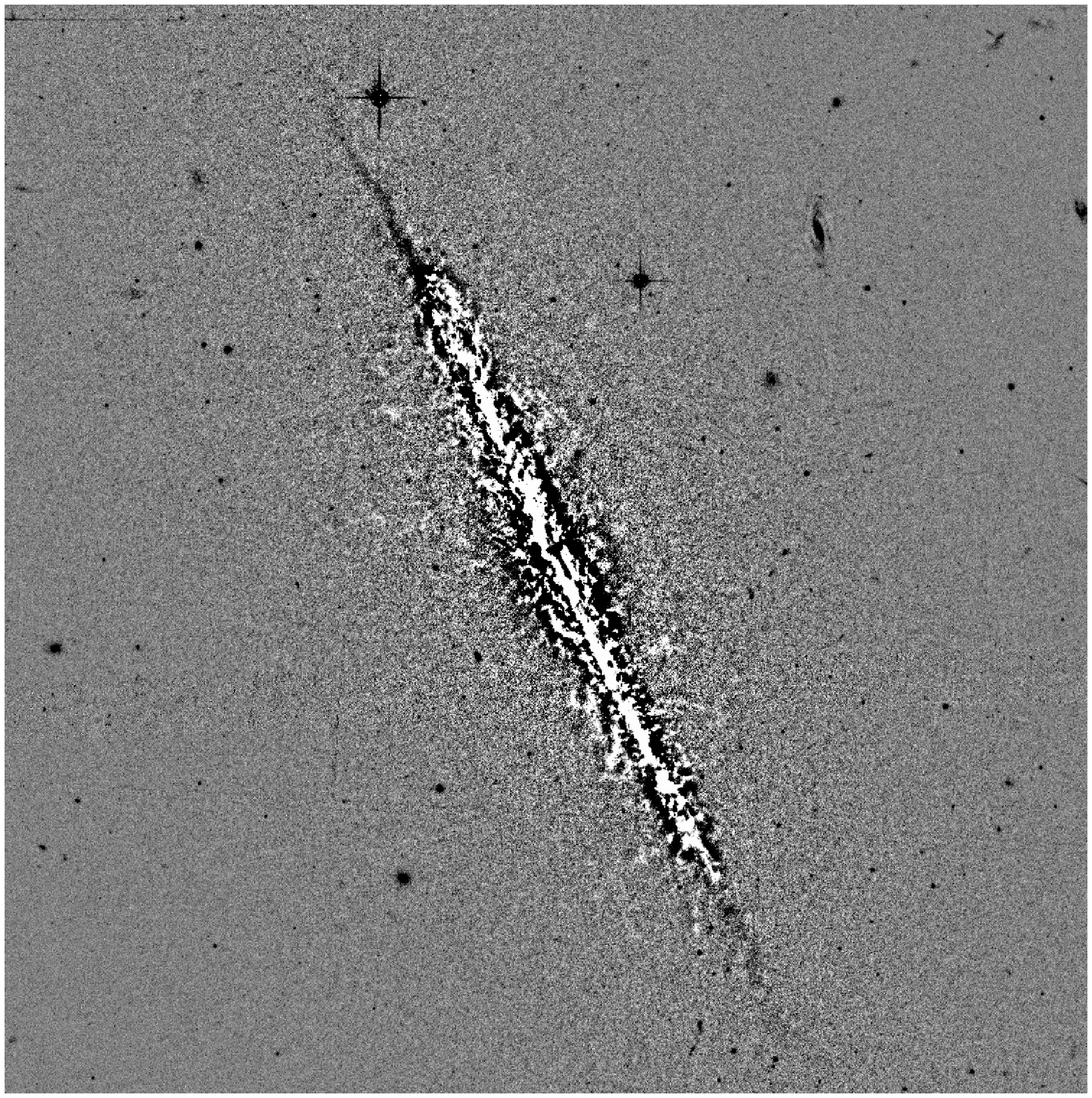}
\caption{Left panel: NGC\,5866 galaxy image from the HST ACS.
The rectangle shows the location of area zoomed in the right
panel. Right panel: A zoom of the disk of dust in NGC\,5866 as seen in
the residual frame.
\label{image}}
\end{figure}

Figure \ref{image} shows the I-band ACS image of
NGC\,5866\footnote{Also visit the web site
http://heritage.stsci.edu/2006/24/index.html for a nice color picture
of the galaxy.} (left panel), and a zoom of the sky+galaxy+large scale
residuals subtracted frame (right panel, ``residual'' frame
hereafter), the nearly edge--on dust disk in the galaxy is easily
recognizable in this second panel.

\section{Globular Cluster selection and photometry}

The residual images were used for the detection and photometry of
point-like and extended sources present in the frames. The source
photometry was obtained independently on all three frames using
SExtractor. After source detection, the output catalogs were matched
using a 0.1$''$ radius.

To derive the aperture correction, standard growth curve analysis
\citep{stetson90} was performed on selected, isolated compact sources.
Extinction corrections \citep[values from][]{sfd98} and aperture
corrections were applied before transforming the ACS magnitudes into
the standard B, V and R magnitudes, using the prescriptions by
\citet{sirianni05}. The transformations from ACS to standard BVRI
system magnitudes were obtained adopting the synthetic transformations
from \citeauthor{sirianni05} We have checked the compatibility of
these magnitudes with the ones derived adopting the empirical
transformations from \citeauthor{sirianni05} As a result, we have
found that the differences between the magnitudes obtained with the
two set of transformations agree on average within 0.02 mag, at least
in the range of colors of GC candidates which will be adopted later
on.

We have chosen to select GC candidates using an approach similar to
\citet[][S06 hereafter]{spitler06}, based on the color, shape and size
of the objects. At the distance of NGC\,5866, the typical GC
half-light radius, $r_{hl}\sim$3 pc, is at the limit of the angular
resolution of the ACS camera (0.05 arcsec/pixel). Under such
condition, the software {\it ishape} \citep{larsen99} can be used to
estimate the intrinsic shape parameters of slightly resolved objects.
To run the $ishape$ software an oversampled PSF of the original image
is needed. Based on their colors, we have selected a sample of star
candidates from each frame and derived the PSF using the photometry
package DAOPHOT, in the IRAF environment. The objects selected for PSF
analysis were a posteriori confirmed to be stars.

In order to save computational time, we have removed from the initial
catalog of matched sources all the sources easily recognizable as
background galaxies. The new catalog was then fed to $ishape$, with
the PSFs, to derive the shape parameters of the listed objects in the
three different passbands. One of the main parameters to run $ishape$
is the convolution kernel, which, once convolved with the PSF, is used
to fit the ellipticity, orientation, FWHM, etc. of the object. After
various tests, we found that the best results were obtained adopting
the analytic King profile, with concentration parameter c=30.

The final shape parameters for each object were then adopted as the
average between the three estimations from the B, V and R
frames. Those objects with half--light radius in one band more than
3--$ \sigma$ different from the V-band $r_{hl}$ value, were rejected
from the catalog. Similarly to S06, we have repeated such rejection
procedure twice on the whole catalog of objects.

Next, we have rejected those object with $ishape$ ellipticities $\geq$
0.5 ($<$ 10 objects at all). As a final selection criteria we have
used the (B-R)$_0$ and (B-V)$_0$ color of the remaining sources. Only
those objects whose color fell within the intervals 0.9$\leq (B-R)_0
\leq$ 1.7, and $0.5 \leq (B-V)_0 \leq 1.1$ were taken as globular
cluster candidates. After these selections, the final catalog of GC
candidates consisted of 109 objects. The list of GC candidates, and
their photometric, spatial and shape properties are reported in Table
\ref{tab_gc}.

Before using these data, it is useful to estimate the amount of
contamination from background galaxies and foreground stars.

Concerning the contamination due to background galaxies, given the
similarity of NGC\,5866 data with those of the Sombrero galaxy by S06,
we adopt the conclusions from these authors. S06 have reduced and
analyzed B and V images from Hubble Deep Field--North of the GOODS
survey, and found that the number of background objects matching the
GC selection criteria is $\sim$ 4\% the final Sombrero GC sample. When
this number is scaled to the imaging area of NGC\,5866 ($\sim$1/6 of
the Sombrero), and to the number of objects classified as GC, we find
that in our case the contamination due to background galaxies is
$\sim$ 6\% the final number of GCs.

The number of galactic foreground stars expected in the color ranges
adopted for GC selection have been calculated using the
\citet{bahcall80} model. A number of approximately $\sim$ 8 stars is
predicted to fall in the range of colors adopted for the GC
selection. We find a similar, but somewhat lower number ($\sim$ 5-6
stars) using the models from \citet{robin03}. These numbers have to be
compared with the number of objects actually matching with the adopted
color ranges, and rejected because of the small half-light radius,
that is 4 objects at all. Given the small number statistics, we
consider the number of contaminating stars from models consistent with
the observed one, thus we can safely assume that the final sample of
GC in Table \ref{tab_gc} is likely free of contamination from
foreground stars\footnote{If the expected and the observed number of
stars are taken as they are, i.e.  without taking into account the
uncertainties, we are lead to conclude that $\sim$ 3\% of the final
GC sample is composed by contaminating stars.}.

\section{Discussion}
\subsection{Color histograms}

\begin{figure}
\epsscale{1.}
\plotone{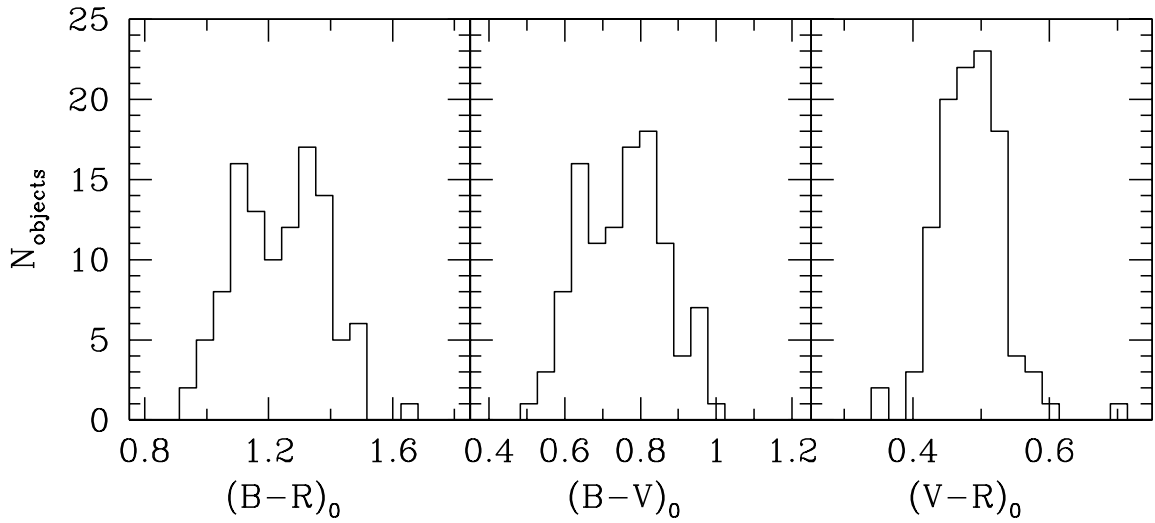}
\caption{Color histograms of the final sample of GC candidates.
\label{colhisto}}
\end{figure}

Figure \ref{colhisto} shows the (B-R)$_0$, (B-V)$_0$ and (V-R)$_0$
color histograms for the 109 GC candidates. To test the bimodality of
the color distribution of the GC sample we use the KMM code
\citep{mclachlan88,ashman94}.  This code evaluates the likelihood
whether or not the color distribution of sample is better represented
by two gaussians or one single gaussian, the positions of the peaks
are evaluated as well.

Running the KMM code on the (B-R)$_0$ color, returns a likelihood of
$\sim$96\% that the GC color distribution is bimodal, with
peaks at (B-R)$_0$=1.12 (blue peak), and (B-R)$_0$=1.35 (red
peak). The GCs are evenly assigned to each group.

An independent KMM run on the (B-V)$_0$ color, returned a slightly
lower but significant bimodality probability (93 \%), and peaks at
(B-V)$_0$=0.67, (B-V)$_0$=0.84 mag.

The narrow range of GCs (V-R)$_0$ color, together with the small
separation between the two possible peaks ($ < 0.1$ mag for the $\sim$
600 GCs of the Sombrero galaxy), and the relatively small number of GC
candidates, hampers any meaningful solution by KMM for this color.

At present, it is still a matter of debate what is the origin of the
bimodality. The recent review from \citet{brodie06} listed some of the
different scenario proposed (major mergers, dissipationless accretion,
hierarchical merging, etc.). In general the photometric data
available, and the few spectroscopic measurements seem to point out
that the primary difference between the two GC subpopulations is due
to metallicity, without the need of invoking age variations. On the
other hand, recently, it has been shown that a nonlinear
color-metallicity relation could explain a bimodal color distribution
without requiring a bimodal metallicity distribution
\citep{richtler06,yoon06}.

\begin{figure}
\epsscale{1.}
\plotone{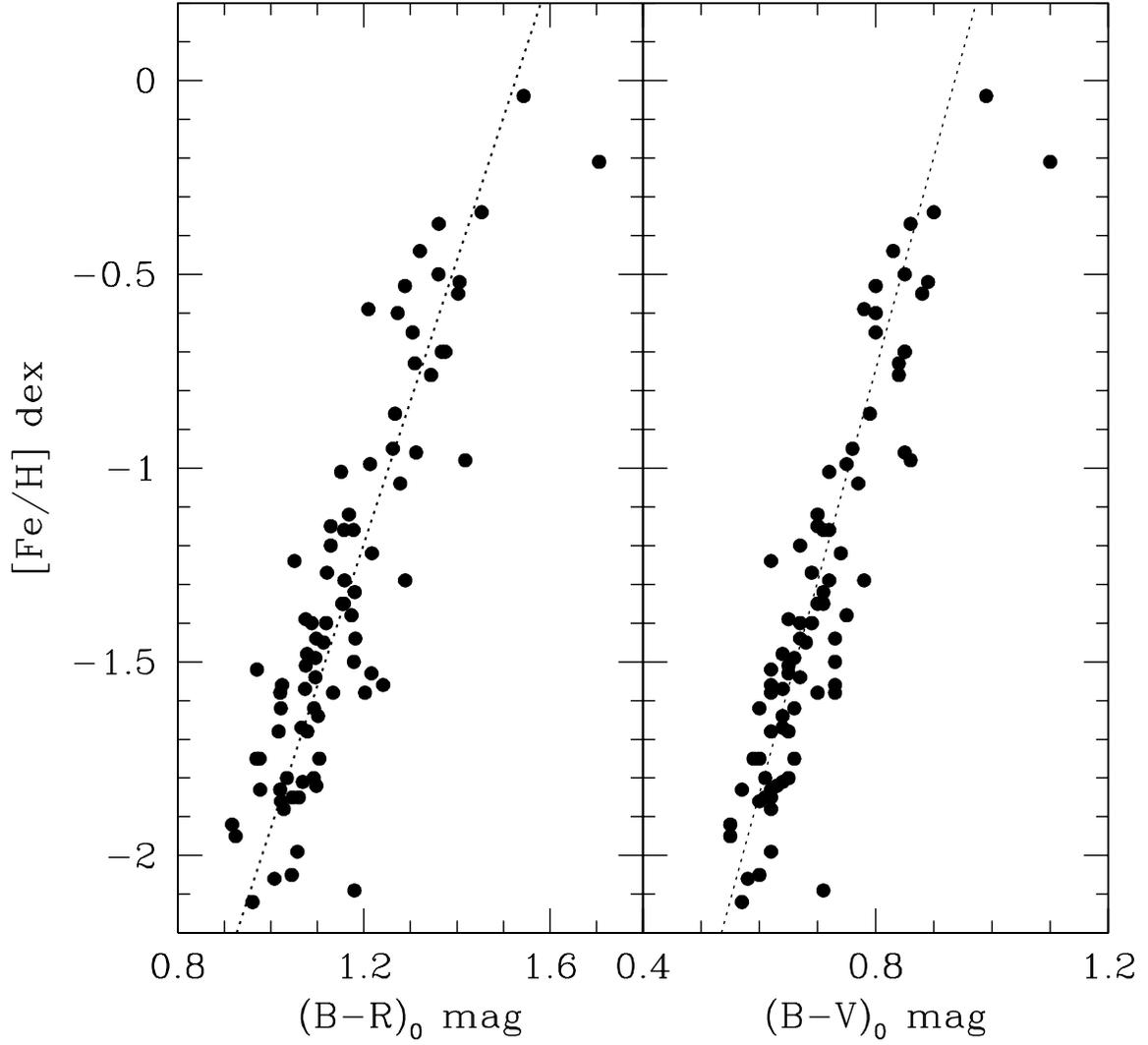}
\caption{Left panel: Galactic GC (B-R)$_0$ color and [Fe/H] data
(full dots) compared with the linear fit to the R05 models (dotted line,
see text).  The fit to the models nicely agrees with observational
data.  Right panel: as left panel, but for (B-V)$_0$ color.
\label{ggc_colfeh}}
\end{figure}

If we assume the picture where the two color peaks are generated by a
bimodal metallicity distribution, the metallicities corresponding to
the peaks can be derived by assuming some [Fe/H] versus color
relations. Adopting the view that GCs are mostly old, and using the
\citet[][R05 hereafter\footnote{The R05 models are available at the
Teramo-SPoT group website: www.oa-teramo.inaf.it/SPoT}]{raimondo05}
models in the metallicity range $-1.8 \leq$ [Fe/H] $\leq$ 0.3, for
ages in the interval 9 $\leq$ t (Gyr) $\leq$ 14, we derive
$[Fe/H]=(3.67 \pm 0.13)(B-R)_0 - (5.60 \pm 0.18)$. We have verified
the reliability of this equation by comparing the theoretical
[Fe/H]$_{th.}$ values predicted for the Galactic GCs, with the
observational [Fe/H]$_{obs.}$ estimations. For this comparison we have
adopted the (B-R)$_0$ and [Fe/H]$_{obs.}$ values from the updated
online \citet{harris96} catalog of Galactic GCs\footnote{The catalog
can be found at the web address:
http://www.physics.mcmaster.ca/$\sim$harris/mwgc.dat}. As a result we
find that the median difference is $\langle
[Fe/H]_{th.}-[Fe/H]_{obs.}\rangle= -0.02 \pm 0.16$ dex, confirming the
reliability of the adopted color-metallicity relation. The good
agreement of the models with Galactic GC data can also be seen in
Figure \ref{ggc_colfeh} (left panel) where we show the fit of the
[Fe/H] versus (B-R)$_0$ from the R05 models, and the Galactic GCs
data.

Using the above [Fe/H] versus (B-R)$_0$ equation, the NGC\,5866 peak
metallicities are [Fe/H]=$-1.48\pm0.23$, and $-$0.64$\pm$0.25 dex for
the blue and red GC peaks, respectively.

We have repeated the above steps using (B-V)$_0$ colors instead of the
(B-R)$_0$. The color-metallicity relation in this case is
$[Fe/H]=(5.48 \pm 0.19)(B-V)_0 - (5.13 \pm 0.16)$  (Fig. \ref{ggc_colfeh},
right panel), which implies [Fe/H]=$-1.49\pm0.20$,
and $-0.54\pm0.23$ dex for the two subpopulations\footnote{Using the
empirical color-metallicity relations derived by S06 from Galactic GC
data, the GC color peaks correspond to [Fe/H]$\sim-1.5$, and
[Fe/H]$\sim-$0.75 dex for the blue and red subpopulations
respectively.}.

These metallicity values can be considered quite average in bimodal GC
systems. In fact, adopting the estimations for other galaxies reported
by \citet{brodie06}, we find on average [Fe/H]=$-1.47\pm0.12$ and
$-0.48\pm0.16$ dex for the sample of normal elliptical and S0 galaxies.

\subsection{Blue tilt}

Inspecting the color magnitude diagram of GCs, it has been shown that
in some giant galaxies a color-luminosity correlation exists for the
blue GCs \citep{harris06}.  The interpretation of this correlation
must be taken carefully, due to the limits of the GC selection
criteria. Moreover, this phenomenon has not been detected in all
galaxies \citep{mieske06a,brodie06}

To explain the existence of the color-luminosity correlation two
different GC formation scenarios have been proposed: the
self-enrichment, and the pre-enrichment one. The differences between
these two scenarios is that in the first one the proto-GCs are
able to trap and absorb the metals formed by stars within the GC;
in the pre-enrichment scenario it is assumed that higher mass proto-GCs
come from intrinsically higher metallicity clouds - see
\citet{strader06} and references therein for more details.

\begin{figure}
\epsscale{1.}
\plotone{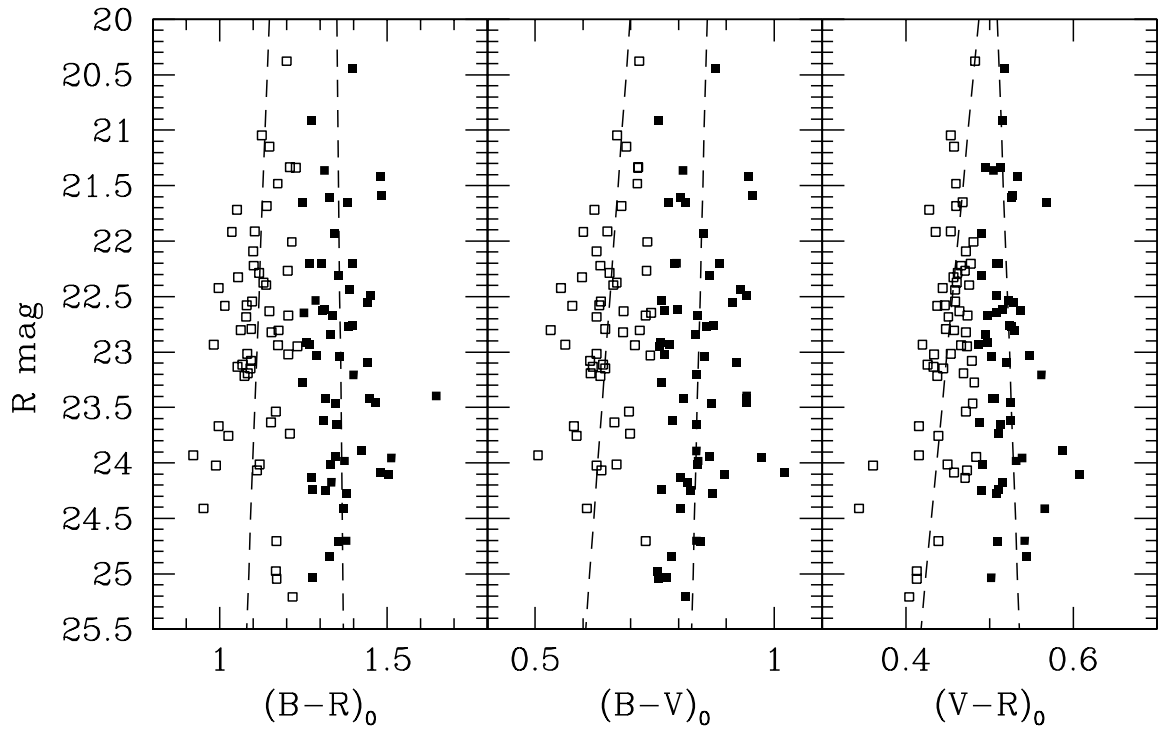}
\caption{Color-Magnitude Diagrams of the final sample of GC. There is
some evidence of the color-magnitude tilt for the blue GC
subpopulation (empty squares).  No significant tilt is found in the
red peak GC candidates (full squares). The dashed lines show the
least-squares fit to the data.
\label{cmd}}
\end{figure}

Figure \ref{cmd} shows the color magnitude diagrams for the NGC\,5866
GCs. For the (B-R)$_0$ and (B-V)$_0$ we independently used the
KMM output classification to separate the GC system into
subpopulations. For the (V-R)$_0$ we used the combined results from
the above colors to separate between the two subpopulations.  It is
worth noting that the sharp separation limits adopted to discriminate
between the two sub-populations introduce a bias in the results, since
real GC subsystems must overlap in some color range. Keeping in mind
this warning, in Figure \ref{cmd} we show the linear least--squares
fits to the single subsystems.

We find a slight color-luminosity correlation for the blue GCs: the
slope of the (B-R)$_0$ color versus R magnitude relation is
$-0.012\pm0.009$. The blue tilt is more evident if (B-V)$_0$ and
(V-R)$_0$ data are taken into account\footnote{For this Figure we have
determined the blue and red peaks of the (V-R)$_0$ color using:
(V-R)$_{0,blue}$=(B-R)$_{0,blue}$-(B-V)$_{0,blue}$=0.45 mag, and
(V-R)$_{0,red}$=(B-R)$_{0,red}$-(B-V)$_{0,red}$=0.51. These numbers
agree with the observed (V-R)$_0$ peaks in the Sombrero (S06).}, since
with slopes $-0.017\pm0.009$, and $-0.013 \pm0.003$, respectively. The
slopes of the color-luminosity relations of the red sub-population are
in all cases consistent with zero.

Using the color metallicity relations reported in the previous section
to convert the tilt in terms of metallicity differences, we find that
the brightest blue GC have on average $\sim$ 0.4 dex higher [Fe/H]
respect to the faintest ones.

It is interesting to note that the study of a sample of bright
ellipticals lead \citet{harris06} to conclude that color-luminosity
relation appears more clearly in galaxies with larger number of
clusters. An analogous result has been found by \citet{mieske06a} by
analyzing the GC properties for a sample of 79 Virgo Cluster galaxies.
A similar conclusion can be drawn by comparing our data with the
results of the Sombrero galaxy by S06. In fact, NGC\,5866 is
morphologically similar to the brighter Sombrero galaxy, but its GC
population within $\sim$1.5 galactic effective radii is $\sim$ 1/5 the
GC population in $\sim$1.5 effective radii of the
Sombrero\footnote{NGC\,5866 images have nearly the same absolute
magnitude completeness limit of the Sombrero images.}, and the
(B-R)$_0$ versus R slope for blue GCs is almost one third the one
derived from the Sombrero galaxy.

\subsection{Spatial properties}
One of the common results concerning the projected radial density of
the GC subpopulations in galaxies, is that they can be fitted by a
power law, with the red subpopulation being more centrally
concentrated than the blue one. More specifically, it has been
shown that the outer parts of GC systems are typically well-fit by
power laws, while the central few kpc, hidden in these NGC\,5866 data,
often show the presence of cores \citep[see, e.g.,][]{ashman98}.

\begin{figure}
\epsscale{1.}
\plotone{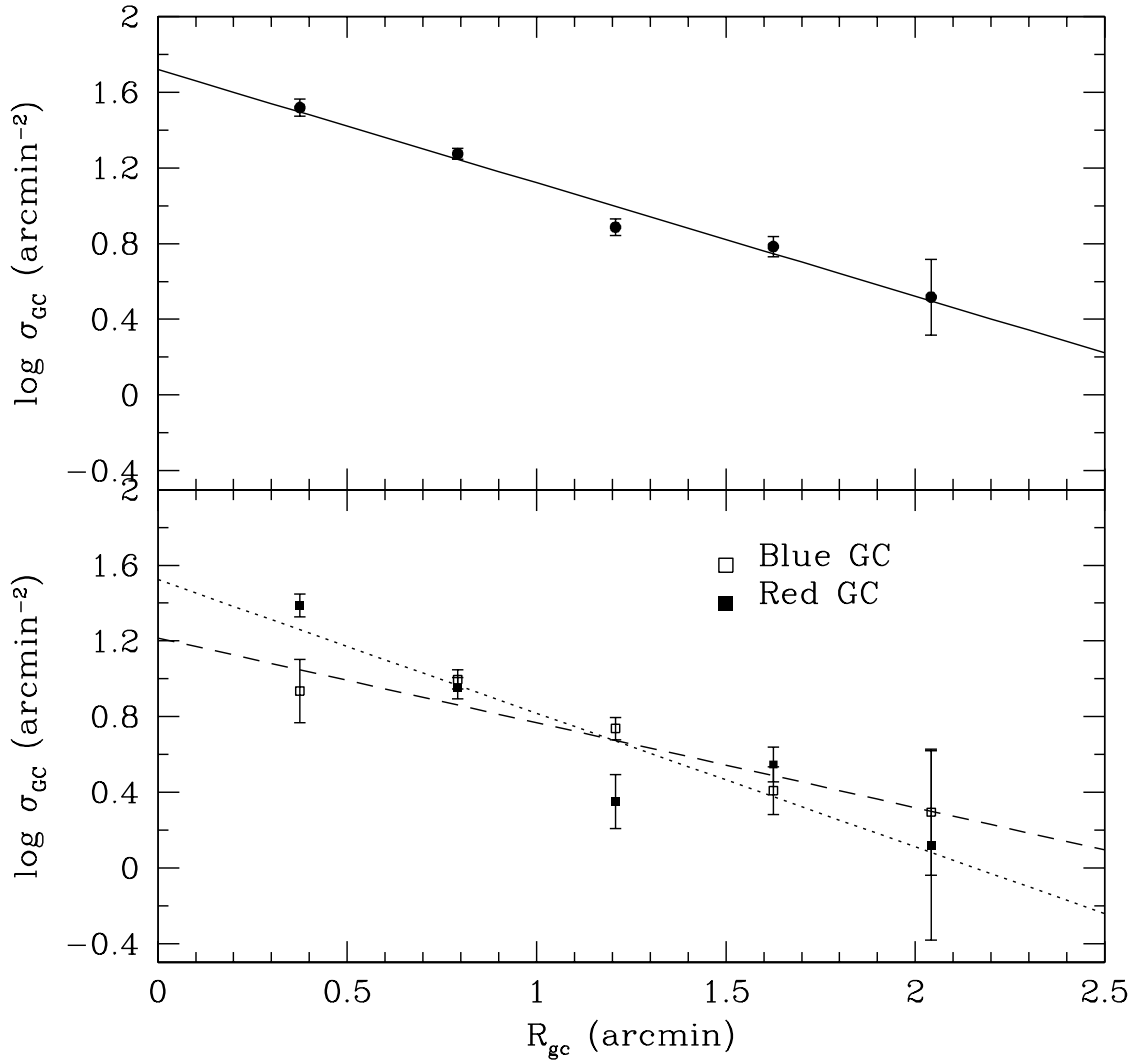}
\caption{Upper panel: The surface density of the entire GC population.
The least-squares profile fit (solid line) shows that the radial
density profile is well reproduced by a power law, $\log \sigma_{GC}
= (-1.3 \pm 0.1) \log R_{gc} + (1.0 \pm 0.1)$. Lower panel: Same as
upper panel but the two subpopulations are shown separately. The
dotted (dashed) line represents a linear fit to the data of red (blue)
GCs. The fitted lines show that the red subpopulation is more
centrally concentrated than the blue one.
\label{radist}}
\end{figure}

Figure \ref{radist} shows the GC surface density in NGC\,5866 versus
the logarithm of galactocentric distance, $\log R_{gc}$. The upper
panel shows the total density profile, while in the lower panel the
profiles of the two GC subpopulations are shown separately. From this
figure it is readily recognizable that $i$) the red subpopulation
appears slightly more centrally concentrated respect to the blue one,
$ii$) the density profiles are well fitted by power laws, with slopes:
$-1.7\pm0.3$, $-0.9\pm0.3$, and $-1.3\pm0.2$ for the red, blue,
and total populations respectively. By comparing the slope of the
total density profile with data from literature we find that NGC\,5866
has an average behavior. In fact, using the compilation by
\citet{kp97}, we find that slopes in the range [$-$2.5,$-$1] are quite
common in galaxies of total magnitude similar to NGC\,5866, i.e.,
$M_{Bt}\sim -20$ mag.

\begin{figure}
\epsscale{1.}
\plotone{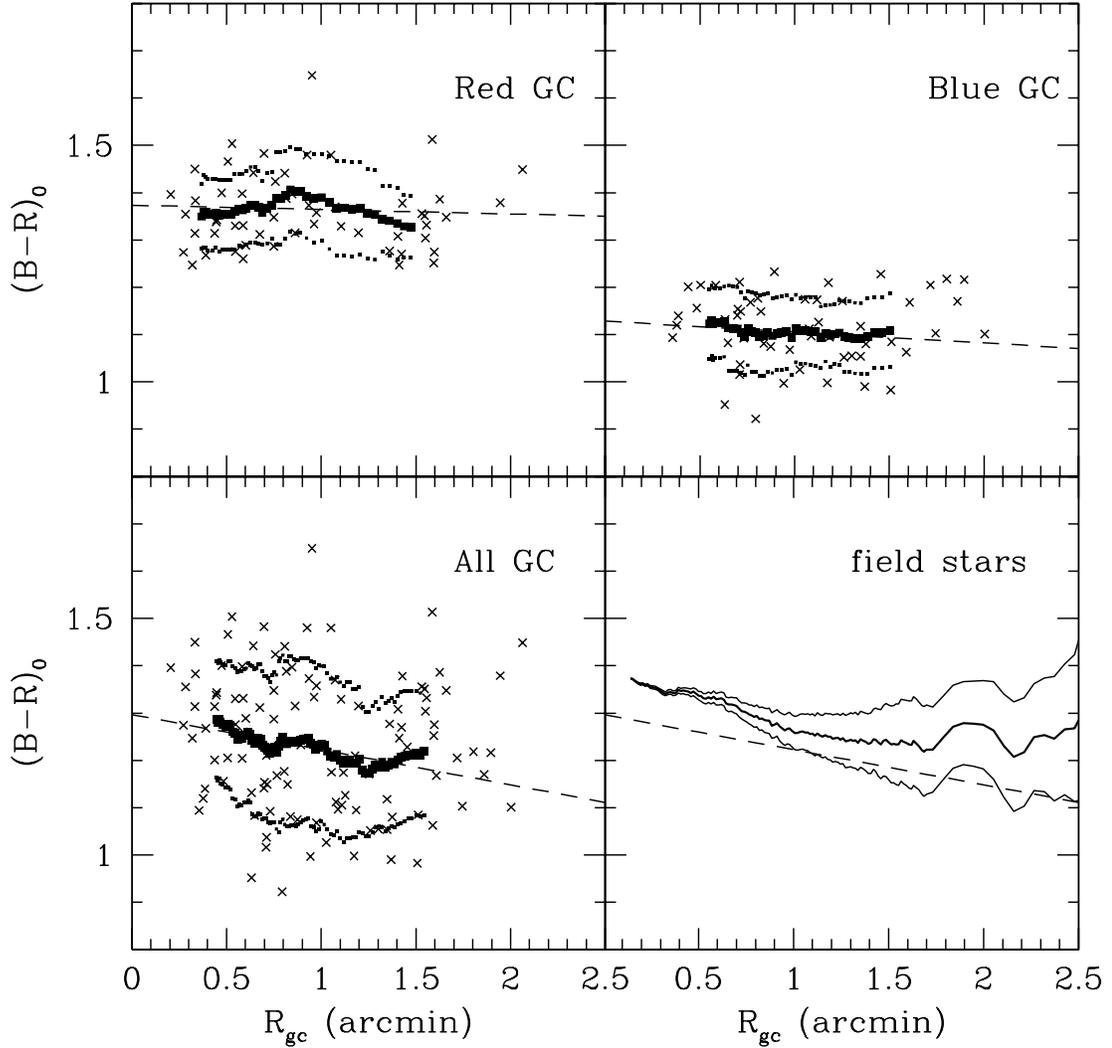}
\caption{GC colors versus the galactocentric distance. The crosses
mark actual GC data. Full big squares correspond to a running average
of data, error bars are shown with small squares. The upper two panels
show that no significant radial color gradient exists if the two
subpopulations are taken separately into account.  If the whole GC
system is considered, a non negligible color gradient is found, with
bluer GC colors in the galaxy outskirts (down left panel). The galaxy
color profile (thick solid line, uncertainties shown with thin lines),
and the least-squares line of the GC system radial profile (dashed
line) are shown in the lower right panel.
\label{colrad}}
\end{figure}

The radial profiles of the (B-R)$_0$ color for the two subpopulations
do not show significant radial gradients (upper panels in Figure
\ref{colrad}). However, due to the different concentrations of the red
and blue GCs shown in Figure \ref{radist}, the total cluster
population shows a clear color gradient (Figure \ref{colrad}, lower
left panel).  In the lower right panel of Figure \ref{colrad} we show
the (B-R)$_0$ color profile of the galaxy, the best fit line to the
total GC color profile is shown, too.  As can be recognized from this
figure, the color profile of field stars closely resembles the color
profile of the total GC system, with a $\sim+0.07$ mag shift,
i.e. $\sim$ 0.3 dex higher metallicity.

These observational properties can be interpreted with a scenario
where the metal poor blue GCs form in the early stages of the galaxy
life, while the red GC component and the field stars are formed later
on in a metal enriched potential well. Or, it could just be a
metallicity gradient for both the GCs and field stars, combined with
the nonlinear color-metallicity relation that originates the observed
GC bimodal color distributions \citep{yoon06,richtler06}.

One other issue in understanding GC system formation and evolution, is
the relation existing between the average sizes of GC subpopulations,
and the variation of half-light radii with galactocentric distances.

Some observational data have shown that the half-light radii of the
red subpopulation are on average 10\% to 20\% smaller than blue
GC. Two different explanations have been given to this phenomenon.
The one proposed by \citet{larsen03} involves a projection effect, due
to the different spatial distribution of red and blue GCs, together
with a strong correlation of the GC size with the galaxy radius. This
model predicts that the size differences should be larger in the inner
galactic regions, and disappear in the outskirts of the galaxy. In the
other scenario, proposed by \citet{jordan04}, mass segregation and
stellar evolution effects are the leading causes of the blue/red GC
size differences. In this scenario little change of relative GC sizes
with galactocentric distances is expected.

\begin{figure}
\epsscale{1.}
\plotone{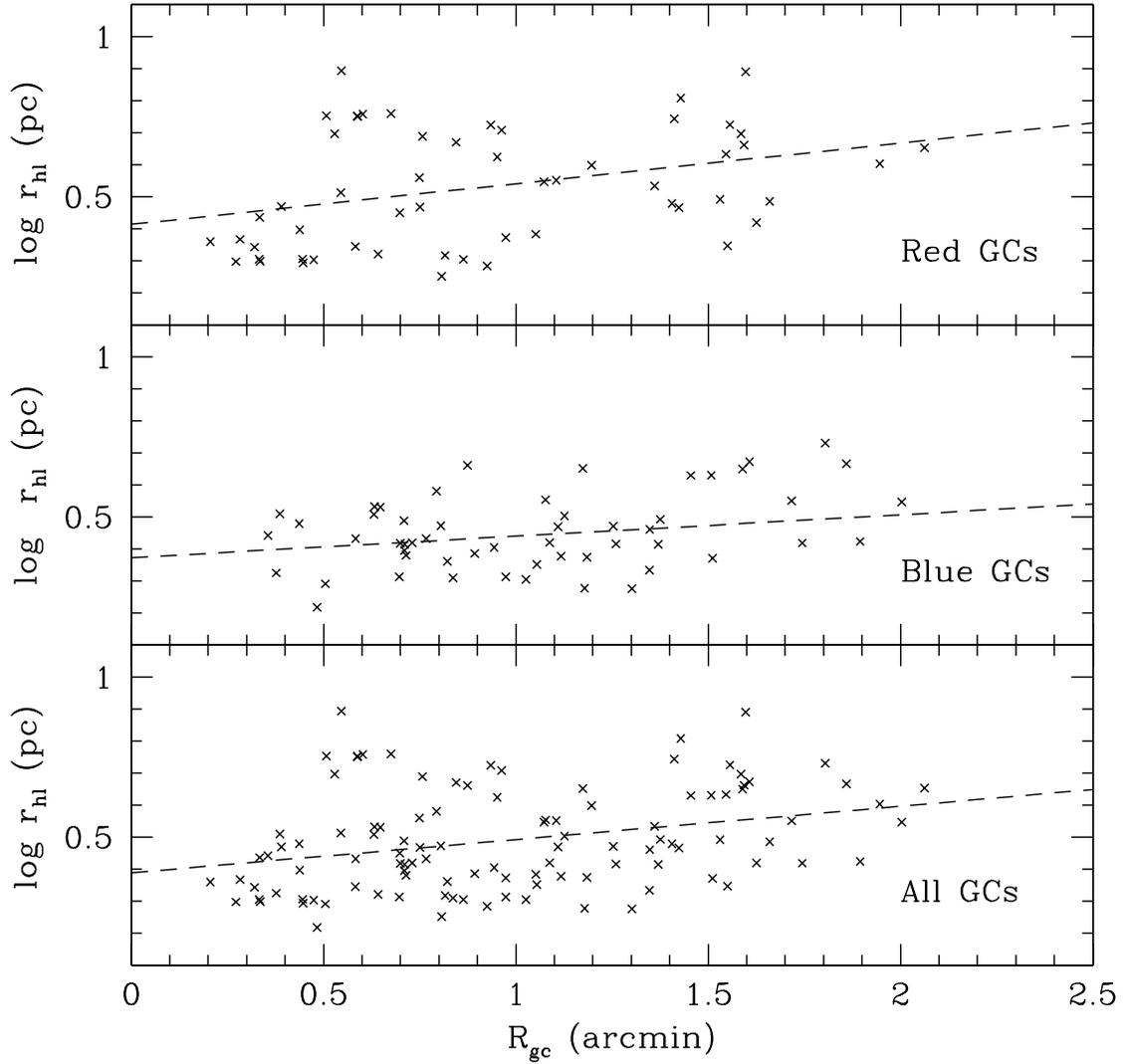}
\caption{Cluster sizes versus the galactocentric distance. The dashed
lines show the least-squares fit to the data.  These data show the
tendency of larger GC radii at larger galactocentric distances.
\label{reffrad}}
\end{figure}

By comparing the average half-light radii of the red and blue GC in
NGC\,5866 we do not find any significant difference between the two
subpopulations. Furthermore, the size versus galactocentric
radius comparison, shown in Figure \ref{reffrad}, reveals a slight
tendency of outer GC to have larger half-light radii. 

Due to the small number of GCs available, the absence of a clear size
difference between the red and blue GCs, and the uncertainties in the
half-light radius versus $R_{gc}$ correlation, these observations
cannot be used to clearly support either the mass segregation or the
projection scenarios.

\subsection{Comparison with models}
\begin{figure}
\epsscale{1.}
\plotone{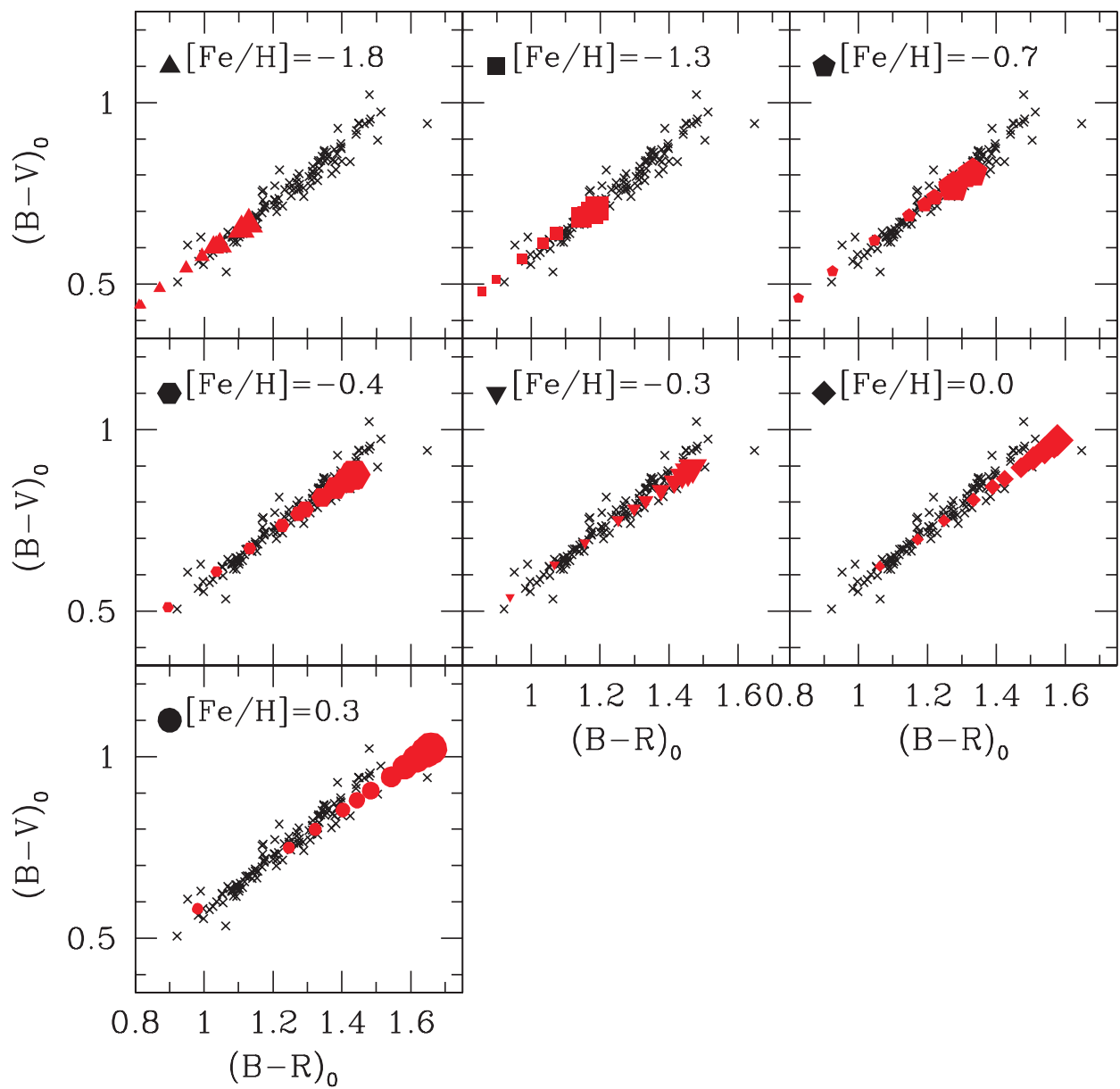}
\caption{Observed GC colors compared with the R05 SSP models for ages
$1 \leq t (Gyr) \leq$ 14 and different metallicities (as
labeled). {\it [See electronic version of the Journal for a color
version of the figure]}
\label{spot}}
\end{figure}

Figure \ref{spot} shows a comparison between the observed colors of
NGC\,5866 GCs and the predictions from R05. Each panel in the figure
shows a Simple Stellar Population (SSP) model with a defined [Fe/H],
and for ages 1 $\leq$ t (Gyr) $\leq$ 14.

At first glance, these panels will give little information about the
age and metallicity properties of the observed GCs, due to the strong
overlap of models with different physical properties. This is not
related to the particular choice of SSP models. In order to reduce
model systematics, we have also considered a data to models comparison
using the \citet{bc03} SSP models. The new set of models essentially
agrees with the R05 one, and the overlap between models at different
[Fe/H] is still present. Moreover, since the R05 models are optimized to
match in detail the observational features of Color Magnitude Diagrams
of Galactic and Magellanic Clouds stellar clusters, in this section we
will take the set of stellar synthesis models from R05 as reference
ones.

If no constraint is put on models, the conclusions that can be drawn
from the comparison shown in Figure \ref{spot} are mainly: $i$) SSP
models, i.e.  single age and single metallicity stellar systems,
reproduce nicely the observational color-color properties of GCs; $ii$)
except one single object, no old super metal-rich (t $>$ 7 Gyr,
[Fe/H]$>$[Fe/H]$_{\sun}$) GC seems to be present; $iii$) the reddest
GCs in the sample [(B-R)$_0\sim$1.6, (B-V)$_0\sim$1] are either old,
t$\sim$ 14 Gyr, solar metallicity objects, or younger, t$\sim$ 5 Gyr,
super-solar metallicity stellar systems.

\begin{figure}
\epsscale{1.}
\plotone{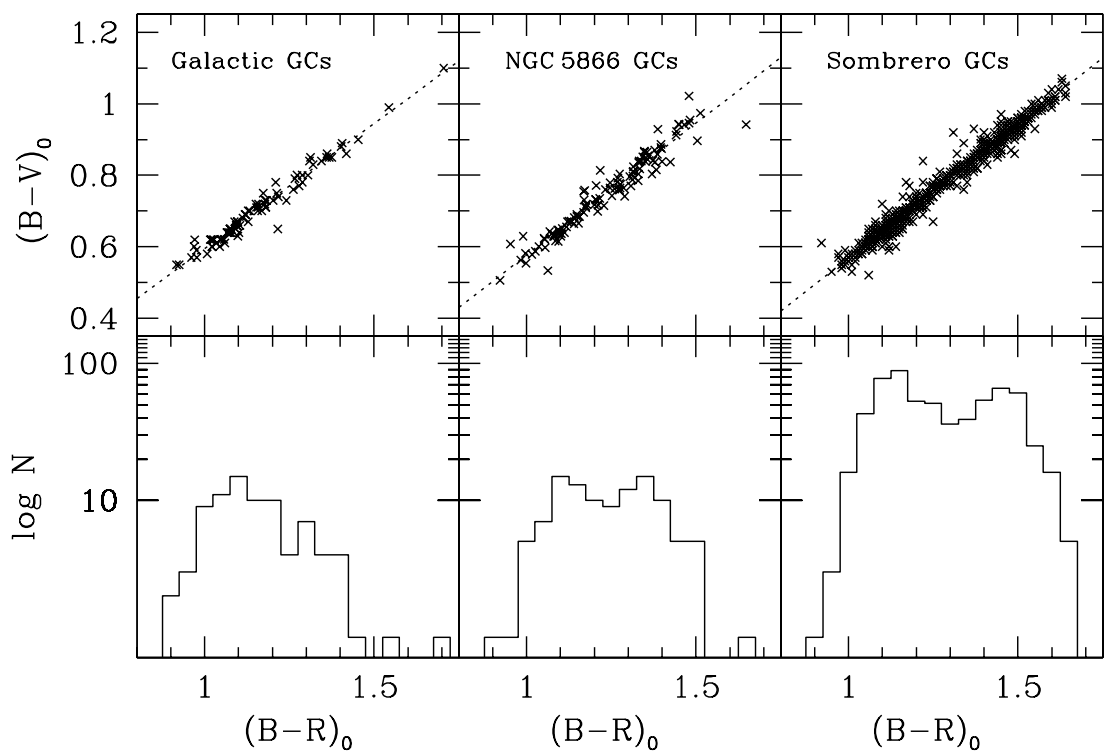}
\caption{Color-color comparisons (upper panels), and (B-R)$_0$ color
histograms (lower panels) for Galactic, NGC\,5866 and Sombrero clusters, as
labeled. The least-squares fit lines are also shown.  The Sombrero GC
system includes a rich component of red, super metal rich globular
clusters which is practically missing in the other two galaxies
system. On the other hand, the blue tails of the GC distributions are
quite similar for the three galaxies.
\label{3gxy}}
\end{figure}

However, if we limit the models to the range of old age models (9
$\leq$ t (Gyr) $\leq$ 14) some stronger constraints on the age and
metallicity of the GC can be obtained. To support this choice, as
mentioned before, we recall that old GC ages have been found in most
of the GC systems in galaxies both with photometric and spectroscopic
data \citep{brodie06}. In addition to this, as shown in Figure
\ref{3gxy} (upper panels), the range of (B-R)$_0$ and (B-V)$_0$
colors for NGC\,5866 and for Galactic GCs appears to be very similar.
If we assume that such color similarity is also associated with common
physical characteristics of the GC systems, than we can adopt for
NGC\,5866 an age range similar to the one of Galactic GC, that is 9
$\lsim$ t (Gyr) $\lsim$ 14 \citep[e.g.][]{deangeli05}.

Based on these considerations, we can go back to the comparisons shown
in Figure \ref{spot}, taking into account only the models in the
selected age interval. With this constraint we find that the blue tail
of GCs has [Fe/H]$\leq-1.8$ dex. If we extrapolate the [Fe/H] versus
(B-R)$_0$ equation derived from R05 models using the same age
limitations adopted here, then the most metal poor GC in NGC\,5866 has
[Fe/H]$\sim-2.3$ dex. On the other hand, inspecting the color properties
of red clusters, the conclusion is that the reddest GCs in NGC\,5866
are old objects, t$\sim$14 Gyr, with nearly solar metallicity.

Before concluding this section, we find interesting to note that if
the same arguments are applied to the GC system in the Sombrero galaxy
(data from S06, shown in the right panels of Figure \ref{3gxy}), the
metallicity range spanned by the GC system is $-2.2\lsim$ [Fe/H]
$\lsim$ 0.6. Thus, while the blue tails of the GC system appear to be
quite similar in the three galaxies, the red tail of the GCs  in
the Sombrero includes also a population of very metal rich clusters.

\subsection{Globular Cluster Luminosity Function}

The review by \citet{richtler03} gives a fairly detailed study on the
use of GCs as distance indicators. By comparing GCLF distances with
those derived using the SBF method, \citeauthor{richtler03} proves
that the TOM is a good distance indicator, although some exceptions
exists, related to a possible contamination from intermediate age GCs,
or young massive clusters.

\begin{figure}
\epsscale{1.}
\plotone{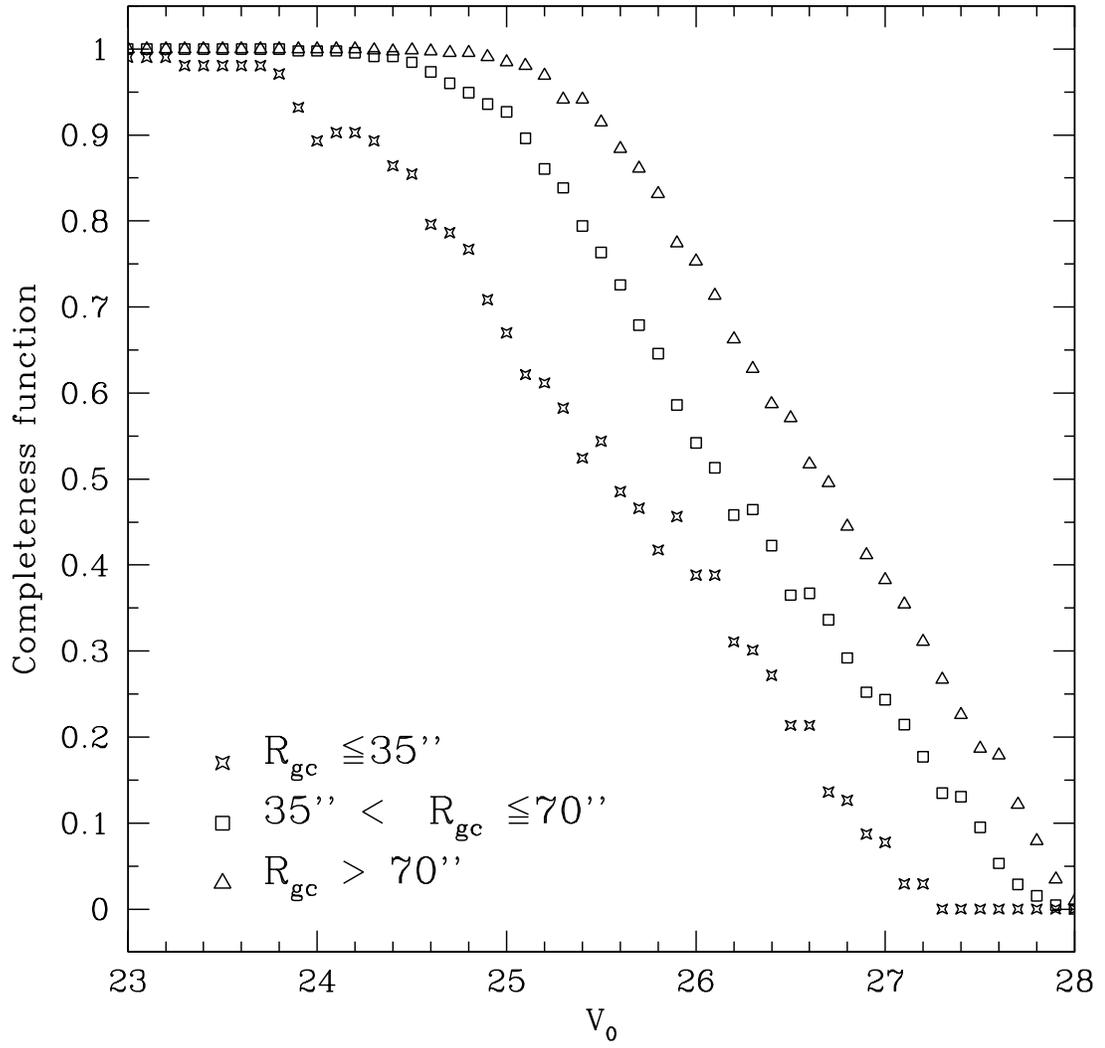}
\caption{V-band completeness functions within three different
galactocentric annuli.
\label{cfv}}
\end{figure}

Figure \ref{cfv} shows the V-band completeness functions for NGC\,5866
derived within three annular regions with radii $R_{gc} (arcsec) \leq
35$, $35 < R_{gc} (arcsec) \leq 70$, and $R_{gc} (arcsec)> 70$. A 95\%
completeness level is reached at V$_0=23.9,~ 24.8,~25.1$ mag for the
three different annuli, respectively. The completeness functions have
been derived via standard artificial star tests, using the DAOPHOT
task $addstar$ to generate the artificial stars, then we ran
SExtractor for the detection of sources, as described in Section 2.

Adopting SBF distances, \citet{richtler03} obtained an absolute TOM
M$_{V,0}=-7.35 \pm 0.25$ mag\footnote{We have corrected the TOM value
reported by \citet{richtler03} using the SBF zeropoint correction as
discussed by \citet{jensen03}.}, which at the distance of NGC\,5866
translates into an apparent TOM V$_0^{TOM}\sim$ 23.46 mag. This means
that the GCLF is fairly complete up to $\sim$0.5, 1.4, and 1.7
magnitudes fainter than the TOM in the three different annular regions
taken into account.

Fitting a gaussian distribution to the GCLF corrected for
incompleteness, gives a TOM V$_0^{TOM}=23.46 \pm 0.06$ mag, with a
dispersion $\sigma=1.13 \pm 0.05$. Adopting the SBF+PNLF average
distance $\mu_0 = 30.75 \pm 0.08$, the absolute TOM is M$_{V,0}^{TOM}=
-7.29 \pm 0.10$ mag, in nice agreement with the value provided by
\citeauthor{richtler03}.

\subsection{Specific Frequency}

Although the field of view of the ACS camera does not sample the
whole NGC\,5866 area, we have obtained an estimation for the Specific
frequency, $S_N$, of GCs over the ACS sampling area. 

The concept of specific frequency, $S_N=N_{GC} \times 10 ^{0.4
(M_V+15)}$, was introduced by \citet{harris81} as a tool to quantify
the amount of stars in GCs with respect to the amount of field stars
in a galaxy. To derive $S_N$ we choose to not do any extrapolation to
derive the total $N_{GC}$ and the total magnitude, $m_V$, to avoid the
uncertainty introduced by the profile extrapolation. Instead, adopting
an approach similar to \citet{blake97}, we adopt the number of GCs
found in the ACS sampling area, after completeness correction, and the
total magnitude obtained from the same masked frame that we used to
detect GCs ($m_V = 10.9 \pm 0.1$ mag). 

As noted by \citet{blake97}, if the GC system follows the same radial
profile of the galaxy light in the outer regions, there should be
little difference between the global $S_N$ and this $S_N$ over a
limited area, referred to as ``metric $S_N$'' by \citeauthor{blake97}

With these assumptions, we derive $S_N= 1.4 \pm 0.3$, which is within
the normal $S_N$ range for galaxies in the same luminosity class of
NGC\,5866 \citep[$S_N$=0.7-7.5,][]{kp97}. We have also estimated the
total $S_N$ extrapolating the total number of GCs. By integrating over
the surface density, corrected for magnitude completeness, we obtained
$N_{GC}^{Total} \sim400$. Adopting a total magnitude $m_V\sim9.8\pm
0.1$ mag from the RC3 catalog, we find $S_N^{Total}\sim 1.6\pm 0.4$,
in nice agreement with the value found over the limited ACS area.

Finally, since the GCs are evenly distributed between the two
subpopulations, we find that the specific frequency for the blue and
red subsystems is $S_N^{blue} \sim S_N^{red} \sim 0.7$.

\section{Conclusions}
We have used archival HST data to study the GC system of NGC\,5866
galaxy in the B, V and R passbands. Using standard color, size and
shape selection criteria, the final list of GC candidates consists of
109 objects. The estimated contamination from background galaxies is
$\sim$ 6 \% of the sample, while essentially no contamination is
expected from foreground stars.

The color distribution of the GC system has a bimodal form. Adopting
the color-metallicity relations from R05 models, we find that the blue
and red peaks correspond to a metallicity [Fe/H]$\sim-1.5$, and $-$0.6
dex, respectively. These metallicity peaks are quite average among the
galaxies with known color bimodality. Similarly, we have found
that the specific frequency of GCs is $S_N=1.4 \pm 0.3$, which is a
normal value for galaxies of the same luminosity class of NGC\,5866.

By inspecting the color magnitude diagram of the two GC subsystems, a
color-luminosity correlation appears in the blue GCs, with brighter
objects being slightly redder than the fainter ones. Interpreting this
feature as a luminosity-metallicity correlation, the bright blue GCs
are found to be 0.4 dex more metal rich than the fainter ones.

Inspecting the spatial properties of the GCs, we find that $i$) the
red subpopulation appears to be more centrally concentrated than the
blue one; $ii$) the radial density of the total GC population, and of
the two subpopulations taken separately, follows a power law radial
profile; and that $iii$) the GCs have larger half-light radii at
larger galactocentric distances.  We do not find any sign of a
significant difference between the average sizes of the
subpopulations.

Taken separately, the two GC subsystems do not show significant color
gradients. However, due to the different radial concentrations, the
total GC system shows a color gradient, very similar to the one of
field stars, except that the GCs are systematically $\sim$0.1 mag
bluer in (B-R)$_0$.  If the GC to field stars color offset is simply
associated to a metallicity difference, then the population of field
stars at a given galactocentric distance is, on average, 0.3 dex more
metal rich of the GC population at the same radius.

A comparison of data with the R05 models has shown that no old, very
metal rich (t$>$9 Gyr, [Fe/H]$>$0) cluster is present. Moreover, if
only models in the age range 9$\leq$ t (Gyr) $\leq$ 14 are taken into
account, then the metallicity range for the GC system in this galaxy
is found to be $-2.3\lesssim$ [Fe/H] $\lesssim 0.0$. When compared to
Galactic and Sombrero data, we find that NGC\,5866 and Milky Way GC
share similar metallicity range, while the Sombrero has a tail of very
metal rich GCs ([Fe/H] $\sim0.6$).

The study of the GC luminosity function has shown that the GC sample
is fairly complete out to $\sim$1.2 mag below the expected TOM. A
gaussian fit to the GCLF provided us V$_0^{TOM}=23.46 \pm 0.06$
mag. Adopting a distance modulus $\mu_0=30.75 \pm 0.08$ from the
weighted average of SBF and PNLF distances, implies
M$_{V,0}^{TOM}=-7.29 \pm 0.10$ mag, in agreement with the mean value
from \citet{richtler03}.

\acknowledgments

We thank the anonymous referee for helping us to improve this paper
with useful suggestions. This work was supported by the NASA grant
AR-10642, and by COFIN 2004 under the scientific project ``Stellar
Evolution'' (P.I.: Massimo Capaccioli).

\clearpage

\begin{figure}
\epsscale{1.}
\plotone{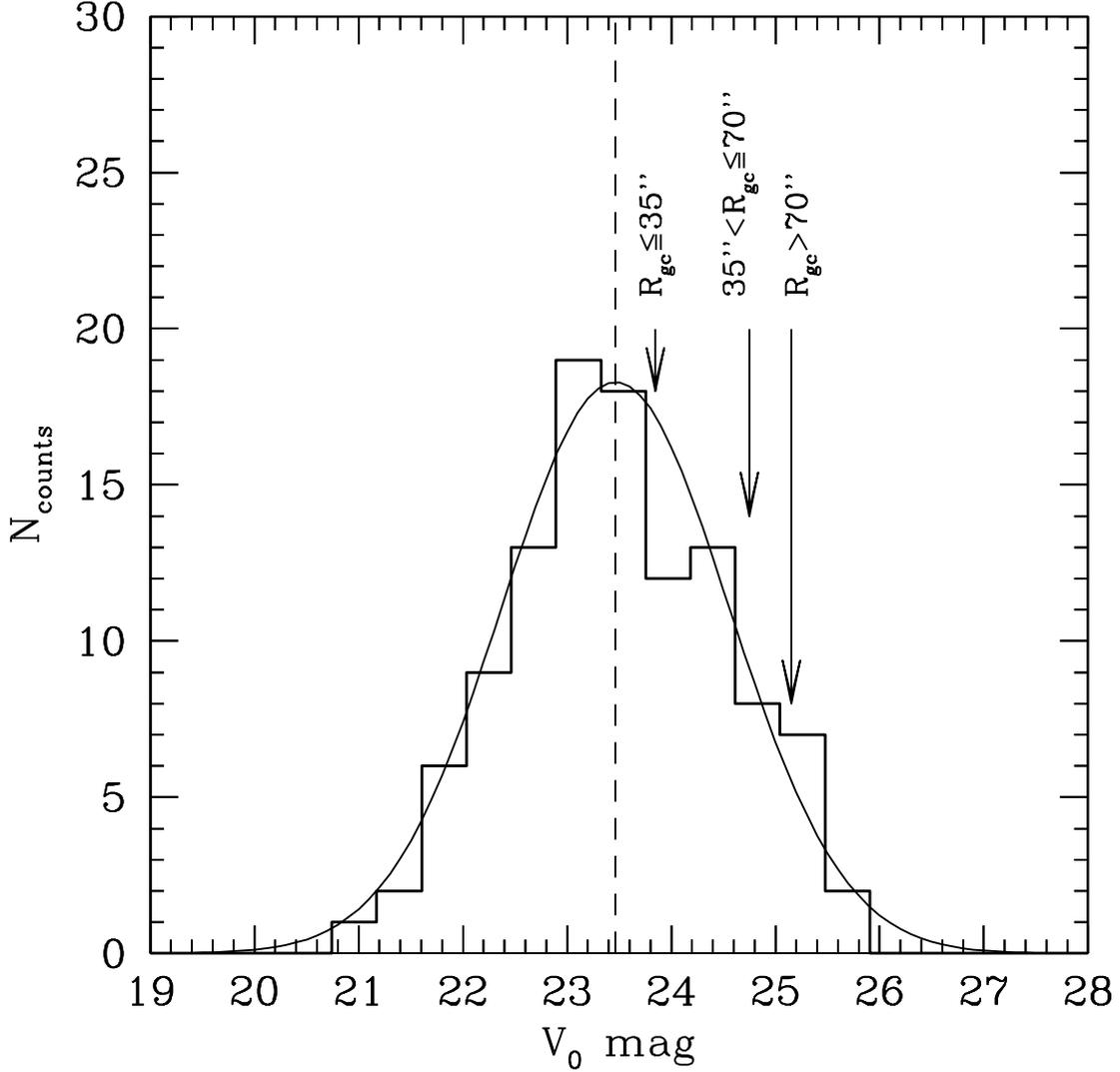}
\caption{The observed GCLF histogram (corrected for
incompleteness) and the best fit gaussian line. The TOM obtained is
V$_0^{TOM} = 23.46 \pm 0.06$ mag (dashed line), dispersion $\sigma =
1.13 \pm 0.05$. Adopting such TOM magnitude we derive the absolute TOM
$M_{V,0}^{TOM}=-7.29 \pm 0.10$ mag, in agreement with the ``universal''
value from \citet{richtler03}. The 95\% completentess magnitude limits
at different galactic annuli are also shown in the figure with
vertical arrows.
\label{gclf}}
\end{figure}

\clearpage

\begin{deluxetable}{ccccccccccc}
\tablecolumns{11}
\tabletypesize{\tiny}
\tablewidth{0pc}
\tablecaption{Globular Clusters in NGC\,5866}
\tablehead{\colhead{R.A.} & \colhead{Decl.} &\colhead{R$_{gc}$} & \colhead{B$_0$} & \colhead{V$_0$}  & \colhead{R$_0$} &
\colhead{(B-R)$_0$} & \colhead{(B-V)$_0$}  & \colhead{(V-R)$_0$}& \colhead{r$_{hl}$} & \colhead{[Fe/H]}}
\startdata
\multicolumn{1}{c}{(J2000)} & \multicolumn{1}{c}{(J2000)} & \multicolumn{1}{c}{arcsec} & \multicolumn{1}{c}{mag} & \multicolumn{1}{c}{mag} & \multicolumn{1}{c}{mag} & \multicolumn{1}{c}{mag} &
\multicolumn{1}{c}{mag} & \multicolumn{1}{c}{mag} & \multicolumn{1}{c}{pc} & \multicolumn{1}{c}{dex} \\
\hline
 15 06 20.43 &  55 46 11.78  & 12.34  &  24.19 $\pm$  0.03  &  23.57 $\pm$  0.03 &   23.13 $\pm$  0.02  &  1.05 &   0.62  &  0.43  &  2.16  &   -1.73 $\pm$  0.20 \\
 15 06 21.23 &  55 46 21.65  & 16.31  &  23.38 $\pm$  0.03  &  22.78 $\pm$  0.02 &   22.33 $\pm$  0.02  &  1.05 &   0.60  &  0.46  &  1.89  &   -1.79 $\pm$  0.18 \\
 15 06 20.06 &  55 45 01.26  & 16.97  &  25.00 $\pm$  0.04  &  24.16 $\pm$  0.03 &   23.65 $\pm$  0.02  &  1.35 &   0.84  &  0.51  &  4.30  &   -0.60 $\pm$  0.22 \\
 15 06 22.81 &  55 47 01.20  & 19.21  &  23.50 $\pm$  0.03  &  22.71 $\pm$  0.02 &   22.20 $\pm$  0.02  &  1.30 &   0.80  &  0.51  &  2.23  &   -0.79 $\pm$  0.20 \\
 15 06 23.30 &  55 46 57.11  & 19.98  &  22.57 $\pm$  0.02  &  21.85 $\pm$  0.02 &   21.34 $\pm$  0.01  &  1.23 &   0.71  &  0.51  &  4.27  &   -1.15 $\pm$  0.18 \\
 15 06 22.41 &  55 45 59.70  & 20.02  &  24.78 $\pm$  0.04  &  24.19 $\pm$  0.03 &   23.75 $\pm$  0.03  &  1.03 &   0.59  &  0.44  &  2.02  &   -1.87 $\pm$  0.22 \\
 15 06 23.15 &  55 46 09.52  & 20.14  &  24.40 $\pm$  0.04  &  23.54 $\pm$  0.03 &   23.04 $\pm$  0.03  &  1.36 &   0.85  &  0.50  &  2.36  &   -0.54 $\pm$  0.23 \\
 15 06 23.84 &  55 46 30.53  & 21.34  &  25.18 $\pm$  0.06  &  24.54 $\pm$  0.05 &   24.07 $\pm$  0.05  &  1.11 &   0.64  &  0.47  &  3.59  &   -1.56 $\pm$  0.28 \\
 15 06 22.32 &  55 45 03.33  & 22.61  &  22.77 $\pm$  0.02  &  22.15 $\pm$  0.02 &   21.72 $\pm$  0.01  &  1.05 &   0.62  &  0.43  &  2.61  &   -1.73 $\pm$  0.18 \\
 15 06 23.97 &  55 46 04.64  & 23.20  &  23.66 $\pm$  0.03  &  23.03 $\pm$  0.03 &   22.58 $\pm$  0.02  &  1.08 &   0.63  &  0.45  &  2.04  &   -1.64 $\pm$  0.20 \\
 15 06 23.84 &  55 45 49.92  & 23.44  &  24.00 $\pm$  0.03  &  23.08 $\pm$  0.02 &   22.56 $\pm$  0.02  &  1.44 &   0.91  &  0.53  &  1.79  &   -0.23 $\pm$  0.22 \\
 15 06 23.34 &  55 45 06.89  & 26.21  &  23.02 $\pm$  0.02  &  22.37 $\pm$  0.02 &   21.91 $\pm$  0.01  &  1.10 &   0.65  &  0.45  &  2.95  &   -1.55 $\pm$  0.18 \\
 15 06 24.73 &  55 46 06.46  & 26.29  &  25.30 $\pm$  0.08  &  24.43 $\pm$  0.06 &   23.95 $\pm$  0.05  &  1.35 &   0.86  &  0.48  &  3.63  &   -0.56 $\pm$  0.34 \\
 15 06 23.02 &  55 44 35.84  & 26.74  &  24.28 $\pm$  0.03  &  23.66 $\pm$  0.02 &   23.19 $\pm$  0.02  &  1.08 &   0.62  &  0.47  &  2.35  &   -1.68 $\pm$  0.19 \\
 15 06 25.16 &  55 46 06.95  & 26.83  &  24.79 $\pm$  0.06  &  24.12 $\pm$  0.05 &   23.63 $\pm$  0.05  &  1.15 &   0.67  &  0.49  &  2.62  &   -1.41 $\pm$  0.28 \\
 15 06 26.45 &  55 47 04.16  & 28.46  &  23.40 $\pm$  0.02  &  22.75 $\pm$  0.02 &   22.29 $\pm$  0.02  &  1.12 &   0.66  &  0.46  &  2.89  &   -1.51 $\pm$  0.18 \\
 15 06 25.51 &  55 46 09.26  & 28.97  &  24.93 $\pm$  0.07  &  24.14 $\pm$  0.06 &   23.62 $\pm$  0.05  &  1.31 &   0.79  &  0.53  &  5.77  &   -0.80 $\pm$  0.33 \\
 15 06 25.09 &  55 45 45.82  & 30.28  &  25.36 $\pm$  0.06  &  24.75 $\pm$  0.05 &   24.41 $\pm$  0.06  &  0.95 &   0.61  &  0.34  &  3.23  &   -1.99 $\pm$  0.30 \\
 15 06 25.95 &  55 46 21.25  & 30.43  &  25.32 $\pm$  0.07  &  24.48 $\pm$  0.05 &   23.89 $\pm$  0.05  &  1.42 &   0.84  &  0.59  &  4.90  &   -0.44 $\pm$  0.33 \\
 15 06 24.71 &  55 45 23.59  & 31.73  &  24.86 $\pm$  0.04  &  24.35 $\pm$  0.03 &   23.93 $\pm$  0.03  &  0.92 &   0.51  &  0.42  &  3.82  &   -2.27 $\pm$  0.22 \\
 15 06 27.61 &  55 47 23.81  & 32.72  &  24.16 $\pm$  0.03  &  23.30 $\pm$  0.02 &   22.77 $\pm$  0.02  &  1.39 &   0.86  &  0.53  &  2.63  &   -0.47 $\pm$  0.21 \\
 15 06 26.05 &  55 46 01.35  & 32.80  &  22.94 $\pm$  0.03  &  22.14 $\pm$  0.02 &   21.61 $\pm$  0.02  &  1.33 &   0.80  &  0.53  &  3.27  &   -0.72 $\pm$  0.21 \\
 15 06 24.40 &  55 44 37.65  & 34.98  &  23.76 $\pm$  0.03  &  23.14 $\pm$  0.02 &   22.68 $\pm$  0.02  &  1.08 &   0.63  &  0.45  &  3.11  &   -1.66 $\pm$  0.19 \\
 15 06 26.79 &  55 46 17.51  & 35.00  &  23.50 $\pm$  0.03  &  22.83 $\pm$  0.03 &   22.37 $\pm$  0.03  &  1.13 &   0.67  &  0.46  &  3.41  &   -1.45 $\pm$  0.21 \\
 15 06 28.11 &  55 47 20.25  & 35.27  &  25.34 $\pm$  0.05  &  24.50 $\pm$  0.03 &   24.01 $\pm$  0.03  &  1.33 &   0.84  &  0.49  &  5.32  &   -0.64 $\pm$  0.25 \\
 15 06 25.97 &  55 45 46.75  & 35.27  &  24.93 $\pm$  0.06  &  23.98 $\pm$  0.04 &   23.46 $\pm$  0.04  &  1.47 &   0.94  &  0.53  &  5.67  &   -0.12 $\pm$  0.28 \\
 15 06 25.72 &  55 45 21.38  & 36.12  &  22.82 $\pm$  0.02  &  22.14 $\pm$  0.02 &   21.68 $\pm$  0.02  &  1.14 &   0.68  &  0.46  &  2.06  &   -1.41 $\pm$  0.18 \\
 15 06 26.25 &  55 45 42.93  & 37.89  &  24.61 $\pm$  0.05  &  23.77 $\pm$  0.04 &   23.21 $\pm$  0.03  &  1.40 &   0.84  &  0.56  &  2.02  &   -0.49 $\pm$  0.25 \\
 15 06 26.96 &  55 46 16.29  & 37.94  &  24.32 $\pm$  0.05  &  23.58 $\pm$  0.04 &   23.03 $\pm$  0.03  &  1.29 &   0.74  &  0.55  &  5.73  &   -0.95 $\pm$  0.25 \\
 15 06 26.02 &  55 46 18.00  & 38.51  &  22.95 $\pm$  0.03  &  22.35 $\pm$  0.02 &   21.92 $\pm$  0.02  &  1.04 &   0.60  &  0.44  &  2.60  &   -1.81 $\pm$  0.19 \\
 15 06 26.51 &  55 45 52.59  & 38.87  &  22.68 $\pm$  0.03  &  21.87 $\pm$  0.02 &   21.36 $\pm$  0.02  &  1.31 &   0.81  &  0.51  &  2.50  &   -0.74 $\pm$  0.20 \\
 15 06 25.81 &  55 45 17.43  & 40.51  &  24.24 $\pm$  0.03  &  23.59 $\pm$  0.03 &   23.15 $\pm$  0.02  &  1.09 &   0.65  &  0.45  &  2.63  &   -1.59 $\pm$  0.20 \\
 15 06 27.12 &  55 46 15.90  & 41.83  &  24.16 $\pm$  0.04  &  23.28 $\pm$  0.03 &   22.76 $\pm$  0.03  &  1.40 &   0.87  &  0.52  &  2.21  &   -0.41 $\pm$  0.24 \\
 15 06 27.78 &  55 46 26.76  & 41.89  &  23.07 $\pm$  0.03  &  22.12 $\pm$  0.02 &   21.59 $\pm$  0.02  &  1.48 &   0.95  &  0.53  &  2.82  &   -0.03 $\pm$  0.21 \\
 15 06 27.84 &  55 46 15.80  & 41.97  &  25.61 $\pm$  0.10  &  24.71 $\pm$  0.07 &   24.10 $\pm$  0.06  &  1.50 &   0.90  &  0.61  &  4.98  &   -0.13 $\pm$  0.41 \\
 15 06 26.59 &  55 45 05.26  & 42.54  &  22.30 $\pm$  0.02  &  21.61 $\pm$  0.02 &   21.15 $\pm$  0.01  &  1.15 &   0.69  &  0.46  &  2.30  &   -1.36 $\pm$  0.18 \\
 15 06 28.82 &  55 46 50.49  & 42.61  &  25.57 $\pm$  0.06  &  24.55 $\pm$  0.04 &   24.09 $\pm$  0.03  &  1.48 &   1.02  &  0.46  &  2.42  &    0.08 $\pm$  0.28 \\
 15 06 27.64 &  55 45 46.80  & 42.66  &  22.18 $\pm$  0.03  &  21.42 $\pm$  0.02 &   20.91 $\pm$  0.02  &  1.27 &   0.76  &  0.52  &  1.99  &   -0.95 $\pm$  0.20 \\
 15 06 28.45 &  55 46 25.36  & 42.86  &  24.10 $\pm$  0.03  &  23.47 $\pm$  0.03 &   23.02 $\pm$  0.03  &  1.08 &   0.63  &  0.45  &  3.40  &   -1.65 $\pm$  0.20 \\
 15 06 28.71 &  55 46 35.43  & 43.84  &  23.98 $\pm$  0.03  &  23.26 $\pm$  0.02 &   22.80 $\pm$  0.02  &  1.18 &   0.72  &  0.46  &  2.97  &   -1.24 $\pm$  0.20 \\
 15 06 26.69 &  55 44 44.60  & 44.97  &  22.17 $\pm$  0.02  &  21.50 $\pm$  0.02 &   21.05 $\pm$  0.01  &  1.13 &   0.67  &  0.45  &  3.19  &   -1.46 $\pm$  0.18 \\
 15 06 27.85 &  55 45 35.12  & 45.04  &  22.90 $\pm$  0.03  &  22.12 $\pm$  0.02 &   21.65 $\pm$  0.02  &  1.25 &   0.78  &  0.47  &  2.21  &   -0.95 $\pm$  0.20 \\
 15 06 27.46 &  55 45 09.04  & 45.43  &  23.60 $\pm$  0.03  &  23.02 $\pm$  0.02 &   22.58 $\pm$  0.02  &  1.02 &   0.58  &  0.44  &  3.08  &   -1.92 $\pm$  0.18 \\
 15 06 27.97 &  55 45 33.02  & 46.01  &  23.94 $\pm$  0.04  &  23.00 $\pm$  0.03 &   22.49 $\pm$  0.03  &  1.45 &   0.94  &  0.51  &  2.02  &   -0.15 $\pm$  0.24 \\
 15 06 26.83 &  55 44 25.46  & 47.61  &  23.47 $\pm$  0.03  &  22.68 $\pm$  0.02 &   22.20 $\pm$  0.02  &  1.27 &   0.79  &  0.48  &  2.93  &   -0.87 $\pm$  0.19 \\
 15 06 29.35 &  55 46 22.64  & 48.30  &  23.47 $\pm$  0.03  &  22.74 $\pm$  0.02 &   22.27 $\pm$  0.02  &  1.20 &   0.73  &  0.47  &  2.71  &   -1.15 $\pm$  0.19 \\
 15 06 29.46 &  55 46 11.11  & 48.43  &  24.20 $\pm$  0.04  &  23.42 $\pm$  0.03 &   22.93 $\pm$  0.03  &  1.27 &   0.78  &  0.49  &  2.96  &   -0.91 $\pm$  0.23 \\
 15 06 29.03 &  55 45 24.94  & 48.95  &  23.54 $\pm$  0.03  &  22.87 $\pm$  0.03 &   22.40 $\pm$  0.02  &  1.14 &   0.66  &  0.48  &  3.24  &   -1.45 $\pm$  0.20 \\
 15 06 30.53 &  55 46 32.97  & 49.34  &  24.70 $\pm$  0.04  &  24.01 $\pm$  0.03 &   23.54 $\pm$  0.03  &  1.17 &   0.70  &  0.47  &  2.71  &   -1.31 $\pm$  0.22 \\
 15 06 29.94 &  55 45 59.62  & 50.20  &  21.84 $\pm$  0.03  &  20.96 $\pm$  0.02 &   20.44 $\pm$  0.02  &  1.40 &   0.88  &  0.52  &  2.29  &   -0.40 $\pm$  0.21 \\
 15 06 31.60 &  55 47 08.10  & 50.70  &  25.01 $\pm$  0.04  &  24.38 $\pm$  0.03 &   24.02 $\pm$  0.03  &  0.99 &   0.63  &  0.36  &  2.60  &   -1.85 $\pm$  0.21 \\
 15 06 29.61 &  55 45 27.53  & 51.80  &  23.03 $\pm$  0.03  &  22.22 $\pm$  0.03 &   21.65 $\pm$  0.02  &  1.38 &   0.81  &  0.57  &  1.99  &   -0.59 $\pm$  0.21 \\
 15 06 30.61 &  55 46 08.54  & 52.50  &  25.13 $\pm$  0.07  &  24.46 $\pm$  0.05 &   24.01 $\pm$  0.05  &  1.12 &   0.67  &  0.45  &  2.12  &   -1.48 $\pm$  0.30 \\
 15 06 31.71 &  55 46 55.76  & 53.60  &  23.42 $\pm$  0.02  &  22.87 $\pm$  0.02 &   22.42 $\pm$  0.02  &  1.00 &   0.55  &  0.44  &  4.49  &   -2.01 $\pm$  0.17 \\
 15 06 29.24 &  55 44 55.99  & 55.55  &  24.74 $\pm$  0.04  &  23.92 $\pm$  0.03 &   23.42 $\pm$  0.03  &  1.31 &   0.81  &  0.50  &  2.02  &   -0.74 $\pm$  0.23 \\
 15 06 30.34 &  55 45 21.76  & 56.10  &  24.01 $\pm$  0.05  &  23.17 $\pm$  0.04 &   22.67 $\pm$  0.04  &  1.34 &   0.84  &  0.50  &  2.02  &   -0.63 $\pm$  0.27 \\
 15 06 30.41 &  55 45 18.26  & 56.60  &  24.23 $\pm$  0.06  &  23.46 $\pm$  0.04 &   23.02 $\pm$  0.04  &  1.21 &   0.77  &  0.43  &  1.96  &   -1.07 $\pm$  0.28 \\
 15 06 31.41 &  55 45 54.54  & 57.10  &  23.67 $\pm$  0.05  &  22.80 $\pm$  0.04 &   22.31 $\pm$  0.04  &  1.35 &   0.86  &  0.49  &  2.33  &   -0.54 $\pm$  0.27 \\
 15 06 31.75 &  55 45 58.62  & 57.75  &  24.17 $\pm$  0.05  &  23.56 $\pm$  0.04 &   23.08 $\pm$  0.04  &  1.09 &   0.61  &  0.48  &  2.77  &   -1.65 $\pm$  0.26 \\
 15 06 31.15 &  55 45 21.98  & 58.45  &  23.98 $\pm$  0.04  &  23.29 $\pm$  0.03 &   22.82 $\pm$  0.03  &  1.16 &   0.68  &  0.47  &  1.65  &   -1.37 $\pm$  0.23 \\
 15 06 31.43 &  55 45 25.88  & 58.45  &  23.28 $\pm$  0.04  &  22.42 $\pm$  0.03 &   21.93 $\pm$  0.03  &  1.34 &   0.85  &  0.49  &  1.97  &   -0.58 $\pm$  0.22 \\
 15 06 31.95 &  55 45 46.99  & 61.60  &  23.93 $\pm$  0.04  &  23.13 $\pm$  0.04 &   22.62 $\pm$  0.03  &  1.31 &   0.80  &  0.52  &  2.74  &   -0.77 $\pm$  0.25 \\
 15 06 31.04 &  55 44 55.54  & 63.15  &  24.18 $\pm$  0.04  &  23.42 $\pm$  0.03 &   22.95 $\pm$  0.02  &  1.23 &   0.76  &  0.47  &  2.44  &   -1.03 $\pm$  0.22 \\
 15 06 30.85 &  55 44 43.28  & 63.30  &  23.64 $\pm$  0.03  &  23.00 $\pm$  0.02 &   22.55 $\pm$  0.02  &  1.10 &   0.64  &  0.46  &  2.63  &   -1.60 $\pm$  0.19 \\
 15 06 32.99 &  55 46 19.41  & 64.35  &  23.78 $\pm$  0.03  &  23.10 $\pm$  0.02 &   22.63 $\pm$  0.02  &  1.15 &   0.69  &  0.46  &  2.40  &   -1.38 $\pm$  0.19 \\
 15 06 32.54 &  55 45 55.30  & 64.60  &  21.58 $\pm$  0.02  &  20.86 $\pm$  0.02 &   20.38 $\pm$  0.01  &  1.20 &   0.72  &  0.48  &  3.02  &   -1.19 $\pm$  0.18 \\
 15 06 30.42 &  55 44 11.48  & 65.30  &  26.14 $\pm$  0.08  &  25.39 $\pm$  0.06 &   24.97 $\pm$  0.05  &  1.17 &   0.76  &  0.41  &  4.71  &   -1.20 $\pm$  0.34 \\
 15 06 32.90 &  55 46 04.62  & 66.30  &  25.52 $\pm$  0.07  &  24.75 $\pm$  0.05 &   24.24 $\pm$  0.05  &  1.28 &   0.77  &  0.51  &  7.83  &   -0.92 $\pm$  0.30 \\
 15 06 31.47 &  55 44 53.38  & 66.55  &  24.66 $\pm$  0.04  &  24.08 $\pm$  0.03 &   23.67 $\pm$  0.03  &  1.00 &   0.58  &  0.42  &  2.54  &   -1.94 $\pm$  0.21 \\
 15 06 32.27 &  55 45 20.77  & 67.05  &  24.18 $\pm$  0.05  &  23.41 $\pm$  0.04 &   22.92 $\pm$  0.03  &  1.26 &   0.76  &  0.50  &  5.66  &   -0.97 $\pm$  0.25 \\
 15 06 32.27 &  55 45 20.77  & 67.60  &  24.18 $\pm$  0.05  &  23.34 $\pm$  0.04 &   22.84 $\pm$  0.03  &  1.33 &   0.83  &  0.50  &  5.66  &   -0.65 $\pm$  0.25 \\
 15 06 32.60 &  55 45 10.58  & 70.45  &  23.82 $\pm$  0.04  &  23.06 $\pm$  0.03 &   22.53 $\pm$  0.03  &  1.29 &   0.76  &  0.52  &  2.94  &   -0.91 $\pm$  0.22 \\
 15 06 32.96 &  55 45 21.88  & 70.70  &  24.54 $\pm$  0.06  &  23.62 $\pm$  0.04 &   23.10 $\pm$  0.04  &  1.44 &   0.92  &  0.52  &  2.10  &   -0.22 $\pm$  0.28 \\
 15 06 32.67 &  55 45 04.23  & 71.05  &  23.60 $\pm$  0.03  &  22.71 $\pm$  0.02 &   22.20 $\pm$  0.02  &  1.40 &   0.89  &  0.51  &  4.69  &   -0.38 $\pm$  0.21 \\
 15 06 31.51 &  55 44 04.28  & 71.85  &  23.32 $\pm$  0.02  &  22.69 $\pm$  0.02 &   22.22 $\pm$  0.02  &  1.10 &   0.64  &  0.47  &  2.63  &   -1.59 $\pm$  0.18 \\
 15 06 34.88 &  55 46 15.02  & 75.20  &  24.29 $\pm$  0.03  &  23.65 $\pm$  0.03 &   23.22 $\pm$  0.02  &  1.07 &   0.64  &  0.44  &  4.59  &   -1.65 $\pm$  0.20 \\
 15 06 33.61 &  55 45 12.48  & 75.60  &  23.83 $\pm$  0.04  &  22.90 $\pm$  0.03 &   22.44 $\pm$  0.02  &  1.39 &   0.93  &  0.46  &  2.08  &   -0.30 $\pm$  0.23 \\
 15 06 33.95 &  55 45 26.45  & 78.15  &  22.54 $\pm$  0.03  &  21.83 $\pm$  0.02 &   21.33 $\pm$  0.02  &  1.21 &   0.72  &  0.49  &  2.48  &   -1.18 $\pm$  0.19 \\
 15 06 36.43 &  55 47 12.95  & 80.85  &  23.87 $\pm$  0.02  &  23.14 $\pm$  0.02 &   22.67 $\pm$  0.02  &  1.21 &   0.73  &  0.47  &  3.56  &   -1.15 $\pm$  0.19 \\
 15 06 33.72 &  55 44 51.40  & 80.90  &  26.17 $\pm$  0.09  &  25.38 $\pm$  0.07 &   24.84 $\pm$  0.07  &  1.33 &   0.79  &  0.54  &  3.56  &   -0.76 $\pm$  0.40 \\
 15 06 36.77 &  55 47 01.61  & 81.65  &  23.90 $\pm$  0.02  &  23.16 $\pm$  0.02 &   22.65 $\pm$  0.02  &  1.25 &   0.74  &  0.51  &  4.60  &   -1.03 $\pm$  0.19 \\
 15 06 36.03 &  55 46 06.86  & 82.20  &  25.51 $\pm$  0.05  &  24.69 $\pm$  0.04 &   24.17 $\pm$  0.04  &  1.33 &   0.82  &  0.52  &  5.11  &   -0.68 $\pm$  0.26 \\
 15 06 35.05 &  55 45 16.93  & 82.55  &  22.90 $\pm$  0.03  &  21.95 $\pm$  0.02 &   21.42 $\pm$  0.02  &  1.48 &   0.95  &  0.53  &  1.92  &   -0.06 $\pm$  0.21 \\
 15 06 35.14 &  55 45 17.05  & 84.35  &  25.36 $\pm$  0.08  &  24.51 $\pm$  0.06 &   23.98 $\pm$  0.05  &  1.37 &   0.84  &  0.53  &  5.30  &   -0.55 $\pm$  0.34 \\
 15 06 34.79 &  55 44 57.06  & 84.70  &  22.65 $\pm$  0.03  &  21.94 $\pm$  0.02 &   21.48 $\pm$  0.02  &  1.17 &   0.71  &  0.46  &  2.39  &   -1.26 $\pm$  0.19 \\
 15 06 36.71 &  55 46 24.84  & 85.40  &  24.95 $\pm$  0.04  &  24.25 $\pm$  0.03 &   23.74 $\pm$  0.03  &  1.21 &   0.70  &  0.51  &  1.90  &   -1.22 $\pm$  0.22 \\
 15 06 35.47 &  55 45 06.86  & 85.70  &  25.78 $\pm$  0.09  &  24.98 $\pm$  0.07 &   24.41 $\pm$  0.06  &  1.37 &   0.80  &  0.57  &  3.53  &   -0.62 $\pm$  0.39 \\
 15 06 36.17 &  55 45 35.06  & 87.30  &  25.05 $\pm$  0.05  &  24.10 $\pm$  0.04 &   23.40 $\pm$  0.03  &  1.65 &   0.94  &  0.71  &  4.22  &    0.27 $\pm$  0.27 \\
 15 06 36.32 &  55 45 34.78  & 90.50  &  24.18 $\pm$  0.03  &  23.54 $\pm$  0.03 &   23.11 $\pm$  0.03  &  1.07 &   0.64  &  0.43  &  2.06  &   -1.65 $\pm$  0.20 \\
 15 06 36.06 &  55 45 15.93  & 90.70  &  24.11 $\pm$  0.04  &  23.40 $\pm$  0.03 &   22.94 $\pm$  0.03  &  1.18 &   0.71  &  0.47  &  2.25  &   -1.27 $\pm$  0.22 \\
 15 06 35.13 &  55 44 24.93  & 91.85  &  25.47 $\pm$  0.05  &  24.49 $\pm$  0.03 &   23.95 $\pm$  0.03  &  1.51 &   0.97  &  0.54  &  4.98  &    0.06 $\pm$  0.27 \\
 15 06 35.30 &  55 44 29.55  & 92.75  &  26.06 $\pm$  0.08  &  25.22 $\pm$  0.06 &   24.71 $\pm$  0.05  &  1.36 &   0.85  &  0.51  &  3.11  &   -0.58 $\pm$  0.34 \\
 15 06 36.26 &  55 44 43.19  & 93.05  &  25.65 $\pm$  0.06  &  24.78 $\pm$  0.04 &   24.27 $\pm$  0.04  &  1.38 &   0.87  &  0.51  &  6.44  &   -0.47 $\pm$  0.30 \\
 15 06 37.02 &  55 45 12.83  & 93.40  &  25.57 $\pm$  0.07  &  24.74 $\pm$  0.05 &   24.25 $\pm$  0.05  &  1.31 &   0.82  &  0.49  &  3.97  &   -0.72 $\pm$  0.32 \\
 15 06 38.97 &  55 46 40.61  & 95.15  &  23.87 $\pm$  0.02  &  23.33 $\pm$  0.02 &   22.80 $\pm$  0.02  &  1.06 &   0.53  &  0.53  &  4.48  &   -1.95 $\pm$  0.18 \\
 15 06 37.47 &  55 45 23.02  & 95.35  &  23.89 $\pm$  0.03  &  23.24 $\pm$  0.02 &   22.79 $\pm$  0.02  &  1.09 &   0.65  &  0.45  &  2.37  &   -1.58 $\pm$  0.20 \\
 15 06 38.47 &  55 45 50.92  & 95.65  &  26.22 $\pm$  0.09  &  25.46 $\pm$  0.07 &   25.05 $\pm$  0.06  &  1.17 &   0.76  &  0.41  &  2.96  &   -1.19 $\pm$  0.38 \\
 15 06 36.56 &  55 43 58.92  & 95.85  &  24.87 $\pm$  0.04  &  23.92 $\pm$  0.03 &   23.42 $\pm$  0.02  &  1.45 &   0.94  &  0.51  &  4.51  &   -0.14 $\pm$  0.23 \\
 15 06 39.13 &  55 45 34.77  & 96.45  &  26.31 $\pm$  0.09  &  25.54 $\pm$  0.07 &   25.03 $\pm$  0.06  &  1.28 &   0.78  &  0.50  &  3.42  &   -0.90 $\pm$  0.39 \\
 15 06 39.01 &  55 45 19.96  & 97.55  &  23.93 $\pm$  0.03  &  23.16 $\pm$  0.02 &   22.62 $\pm$  0.02  &  1.31 &   0.77  &  0.54  &  3.02  &   -0.85 $\pm$  0.20 \\
 15 06 39.55 &  55 45 38.10  & 99.60  &  24.52 $\pm$  0.03  &  23.76 $\pm$  0.02 &   23.27 $\pm$  0.02  &  1.25 &   0.77  &  0.48  &  5.55  &   -0.98 $\pm$  0.21 \\
 15 06 39.04 &  55 44 48.27  & 103.00 &  24.81 $\pm$  0.04  &  23.94 $\pm$  0.03 &   23.46 $\pm$  0.03  &  1.35 &   0.87  &  0.48  &  3.06  &   -0.53 $\pm$  0.24 \\
 15 06 39.68 &  55 45 17.58  & 104.70 &  23.92 $\pm$  0.03  &  23.35 $\pm$  0.02 &   22.93 $\pm$  0.02  &  0.98 &   0.56  &  0.42  &  4.27  &   -2.02 $\pm$  0.18 \\
 15 06 39.77 &  55 45 05.47  & 108.25 &  25.41 $\pm$  0.05  &  24.61 $\pm$  0.04 &   24.14 $\pm$  0.04  &  1.27 &   0.80  &  0.47  &  7.78  &   -0.84 $\pm$  0.27 \\
 15 06 41.73 &  55 46 22.37  & 111.55 &  26.43 $\pm$  0.09  &  25.61 $\pm$  0.07 &   25.21 $\pm$  0.06  &  1.22 &   0.81  &  0.40  &  5.39  &   -0.97 $\pm$  0.38 \\
 15 06 41.74 &  55 45 04.14  & 113.70 &  25.87 $\pm$  0.07  &  25.14 $\pm$  0.05 &   24.70 $\pm$  0.05  &  1.17 &   0.73  &  0.44  &  4.64  &   -1.24 $\pm$  0.31 \\
 15 06 42.57 &  55 45 17.59  & 116.75 &  23.22 $\pm$  0.02  &  22.49 $\pm$  0.02 &   22.01 $\pm$  0.02  &  1.22 &   0.73  &  0.48  &  2.66  &   -1.12 $\pm$  0.19 \\
 15 06 42.26 &  55 45 01.19  & 120.20 &  26.08 $\pm$  0.08  &  25.25 $\pm$  0.06 &   24.70 $\pm$  0.05  &  1.38 &   0.84  &  0.54  &  4.02  &   -0.54 $\pm$  0.34 \\
 15 06 43.79 &  55 45 54.55  & 123.75 &  23.19 $\pm$  0.04  &  22.56 $\pm$  0.03 &   22.09 $\pm$  0.03  &  1.10 &   0.63  &  0.47  &  3.53  &   -1.61 $\pm$  0.22 \\
\enddata
\label{tab_gc}
\end{deluxetable}

\clearpage

\bibliographystyle{apj}
\bibliography{cantiello_jun07}

\end{document}